\documentclass[twocolumn,superscriptaddress]{revtex4}

\usepackage{doi}
\usepackage{hyperref}
\hypersetup{
  colorlinks=true,        
  linkcolor=blue,         
  citecolor=cyan,         
}

\usepackage{graphicx}
\usepackage{dcolumn}
\usepackage{bm}
\usepackage{color}
\usepackage{enumitem}
\usepackage{amsmath}
\usepackage{subcaption}
\usepackage{amssymb}
\usepackage{orcidlink}
\usepackage{soul}
\usepackage{xcolor}
\usepackage{float}

\begin{document}

\title{Gravitational wave signatures of periodic orbits around a Schwarzschild-like black hole}

\author{Mirzabek Alloqulov,\orcidlink{0000-0001-5337-7117}}
\email{malloqulov@gmail.com}

\affiliation{School of Physics, Harbin Institute of Technology, Harbin 150001, People’s Republic of China}

\affiliation{University of Tashkent for Applied Sciences, Str. Gavhar 1, Tashkent 100149, Uzbekistan}

\affiliation{Department of Physics, New Uzbekistan University, Movarounnahr str. 1, Tashkent 100000, Uzbekistan}

\author{Sanjar Shaymatov, \orcidlink{0000-0002-5229-7657}}
\email{sanjar@astrin.uz}
\affiliation{Institute of Fundamental and Applied Research, National Research University TIIAME, Kori Niyoziy 39, Tashkent 100000, Uzbekistan}
\affiliation{National University of Uzbekistan, Tashkent, 100174, Uzbekistan}

\author{Bobomurat Ahmedov,\orcidlink{0000-0002-1232-610X}}
\email{ahmedov@astrin.uz}

\affiliation{Institute for Advanced Studies, New Uzbekistan University, Movarounnahr str. 1, Tashkent 100000, Uzbekistan}
\affiliation{School of Physics, Harbin Institute of Technology, Harbin 150001, People’s Republic of China}
\affiliation{Institute of Theoretical Physics, National University of Uzbekistan, Tashkent 100174, Uzbekistan}

\author{Chengxun Yuan,\orcidlink{0000-0002-2308-6703}}
\email{yuancx@hit.edu.cn}

\affiliation{School of Physics, Harbin Institute of Technology, Harbin 150001, People’s Republic of China}

\date{\today}

\begin{abstract}
We investigate how deviations from the Schwarzschild geometry affect the motion of massive test particles and the gravitational radiation associated with their periodic trajectories around a new improved Schwarzschild black hole. Periodic configurations are determined by examining the commensurability of the radial and azimuthal orbital frequencies. For representative trajectories, we numerically reconstruct the particle motion and calculate corresponding gravitational wave polarizations using the symmetric trace-free mass-quadrupole approach. By comparing these signals with their Schwarzschild counterparts, we assess how the modified spacetime geometry is encoded in the radiation. Our analysis shows that the parameters $\xi$ and $\gamma$ produce modest changes in the orbital configurations and their associated waveforms, providing a means of investigating the dynamical signatures of this BH model.
\end{abstract}

\maketitle

\section{Introduction}

The motion of compact objects in the strong gravitational field of a black hole (BH) provides a way to investigate how the geometry of spacetime governs observable astrophysical phenomena. In particular, the trajectories of small bodies orbiting supermassive BHs encode information about the gravitational field through their orbital frequencies. The observations of gravitational waves (GWs) from compact binaries~\cite{Abbott_2016} and the imaging of BH shadows~\cite{Akiyama19L1, Akiyama19L6, Event} have made it possible to investigate these systems through both gravitational and electromagnetic measurements. An important question arising from these developments is whether observations can distinguish the standard BH solutions of general relativity (GR) from those modified by additional physical effects. Addressing this question requires a detailed understanding of how changes in the spacetime metric influence particle motion and the radiation emitted by orbiting bodies.

One motivation for considering modified BH geometries comes from the difficulty of reconciling classical gravity with the physics expected at extremely high energies. The formation of spacetime singularities in gravitational collapse indicates that the classical description may require modification in regions where curvature becomes sufficiently large~\cite{Penrose1965}. Several approaches to quantum gravity seek to address this limitation, including asymptotic safety, in which the gravitational interaction is described by scale-dependent couplings governed by renormalization group flow~\cite{Reuter1998, Niedermaier2006, Reuter2019}. In this framework, the effective gravitational coupling need not remain constant across all length scales. A renormalization group improvement of a classical BH solution incorporates this scale dependence into its metric and provides a phenomenological setting for exploring possible quantum corrections to gravitational dynamics~\cite{Bonanno2000, Bonanno2006, Falls2018}. Such an effective description makes it possible to examine whether modifications that preserve the familiar large-distance behavior nevertheless produce measurable changes in the strong-field region. The present study considers a new improved Schwarzschild BH geometry described by two additional parameters $\xi$ and $\gamma$, which are the cutoff scale and interpolation parameters~\cite{ALENCAR2026170756}.

Among bound trajectories, periodic orbits are especially useful because their radial and azimuthal motions repeat in a definite pattern~\cite{Levin_2008}. Their closure is associated with a rational relation between the fundamental frequencies, allowing the trajectories to be classified according to their resonant structure. In the strong-field regime, these orbits can depart substantially from simple elliptical motion and develop intricate patterns involving extended radial excursions and repeated loops near periapsis. The zoom, whirl, and vertex numbers provide a systematic description of these configurations~\cite{Levin_2009}. This classification makes it possible to compare orbits with different geometric structures and to determine how the underlying gravitational field changes their morphology~\cite{Levin2010, Mustapha2020}. These orbits are also useful for calculating gravitational radiation from extreme mass ratio inspiral (EMRI) systems~\cite{Hughes01EMRI, Amaro-Seoane18LRR, Babak17PRD}. EMRIs consist of a stellar-mass compact object orbiting a supermassive BH and are among the principal targets of future space-borne GW observatories, including Laser Interferometer Space Antenna (LISA)~\cite{Amaro-Seoane2017LISA}, Taiji~\cite{HuTaiji} and TianQin~\cite{Gong2021}. Their long orbital evolution offers the possibility of probing the central spacetime through detailed measurements of the emitted signal. Various works investigated the gravitational wave signatures from periodic orbits by considering different spacetime geometries~\cite{Zhou:2020zys, Zahra:2025tdo, Deng:2025wzz, Li:2025eln, Deng:2020hxw, Wei:2025qlh, Alloqulov:2025bxh, Wang:2025wob, Healy:2009zm, Zhang:2022zox, Pugliese:2013xfa, Wei:2019zdf, Azreg-Ainou:2020bfl, Gao:2020wjz, Lin:2021noq, Gao:2021arw, Lin:2022wda, Zhang:2022psr, Habibina:2022ztd, Haroon:2025rzx, Lin:2023eyd, Wang:2022tfo, Chan:2025ocy, Lin:2022llz, Yao:2023ziq, Lin:2023rmo, Liu:2018vea, Alloqulov:2025ucf, Choudhury:2025qsh, Li:2025sfe, Zare:2025aek, Chen:2025aqh, Lu:2025cxx, Jiang:2024cpe, Junior:2024tmi, Shabbir:2025kqh, Yang:2024lmj, Wang:2025hla, QiQi:2024dwc, Yang:2024cnd, Li:2024tld, li2025periodicorbitsgravitationalwaves, deng2025gravitationalwaveformsperiodicorbits, Sharipov:2025yfw, Sharipov:2026CPC, Ahmed:2025azu, Gong:2025mne, Glampedakis2002, HEIDARI2026170773, gogoi2026equatorialperiodicorbitsgravitational, shahzad2026periodicorbitsgravitationalwavesignatures, https://doi.org/10.1002/andp.70303}. 

In this paper, we explore the orbital dynamics and GW radiation from the periodic motion around the new improved Schwarzschild BH. We begin by deriving the effective potential for massive particles and examining its behavior as the orbital angular momentum and spacetime parameters vary. The marginally bound orbits (MBOs) and innermost stable circular orbits (ISCOs) are then investigated numerically to establish the relevant boundaries for bound motion. We subsequently construct periodic trajectories characterized by different combinations of zoom, whirl and vertex numbers, using the rational frequency relation. Finally, we calculate the two GW polarizations for representative periodic orbits and compare the resulting signals with those obtained in the Schwarzschild limit.

The article is organized as follows. Section~\ref{sec2} introduces the BH metric and develops the equations governing massive particle motion, including the conditions used to determine the MBO and ISCO. Section~\ref{sec3} presents the periodic orbit analysis and discusses the dependence of the trajectories on the orbital classification numbers and spacetime parameters. Section~\ref{sec4} describes the calculations of gravitational waveforms and compares the signals associated with different orbital configurations. The main results and their implications are summarized in Section~\ref{sum}. The work adopts the metric signature $(-, +, +, +)$ and geometric units where $G=c=1$.

\section{Spacetime}\label{sec2}

\begin{figure*}
\includegraphics[scale=0.3]{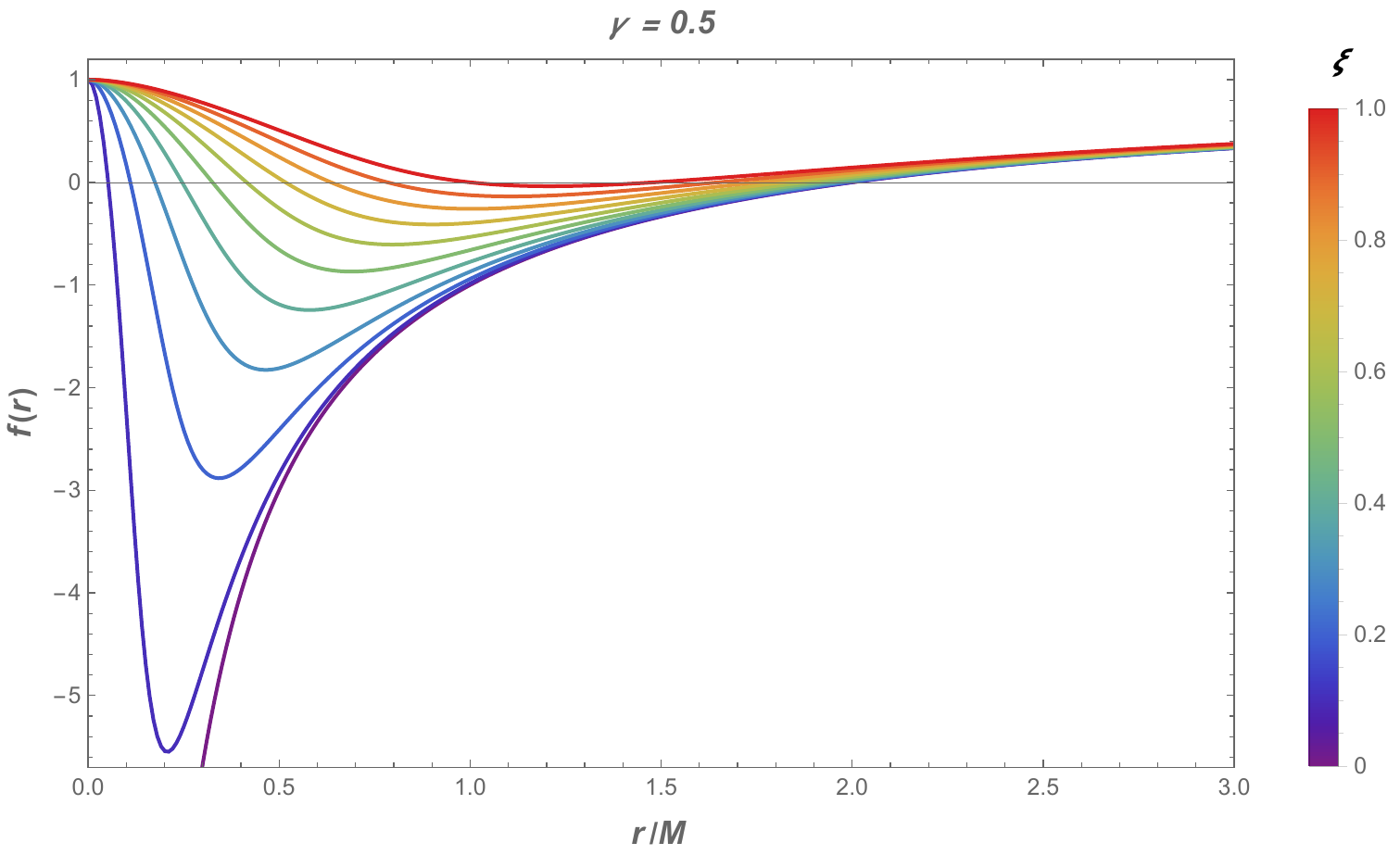}
\includegraphics[scale=0.3]{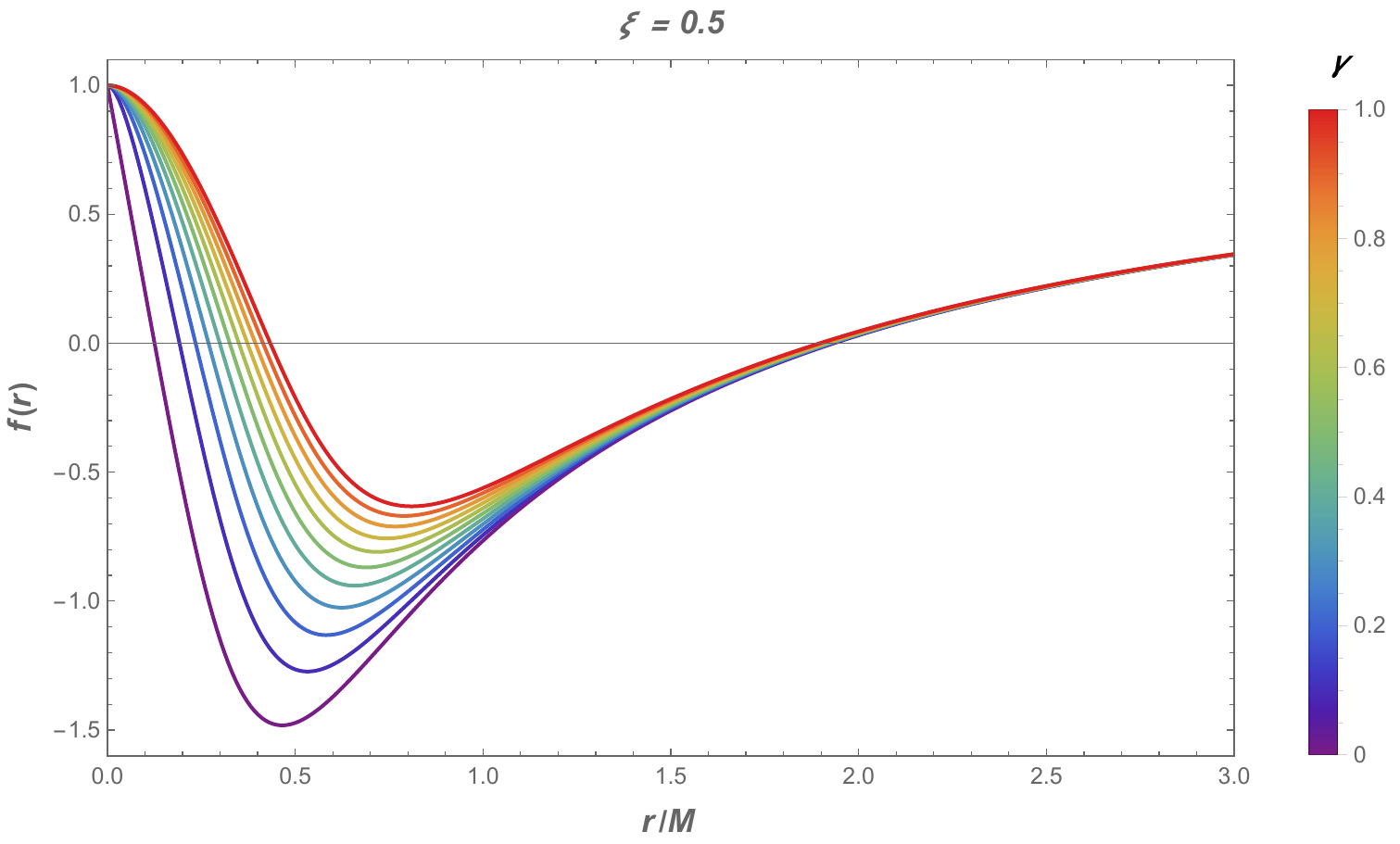}
\caption{The radial dependence of the metric function $f(r)$ for the different values of the $\xi$ (left panel) and $\gamma$ (right panel) parameters. We set $\gamma=0.5$ and $\xi=0.5$ for the left and right panels, respectively.}
\label{fig:metric}
\end{figure*}
\begin{figure*}
\includegraphics[scale=0.4]{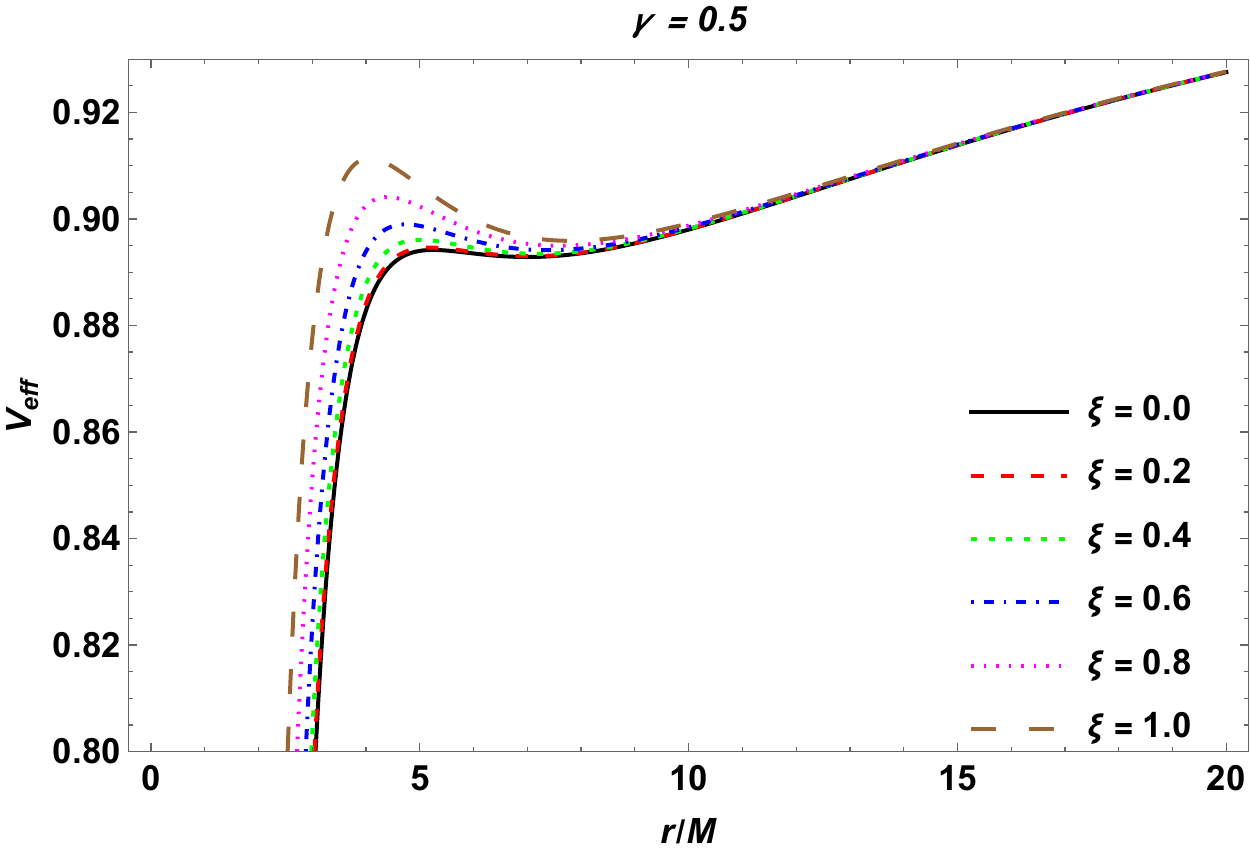}
\includegraphics[scale=0.4]{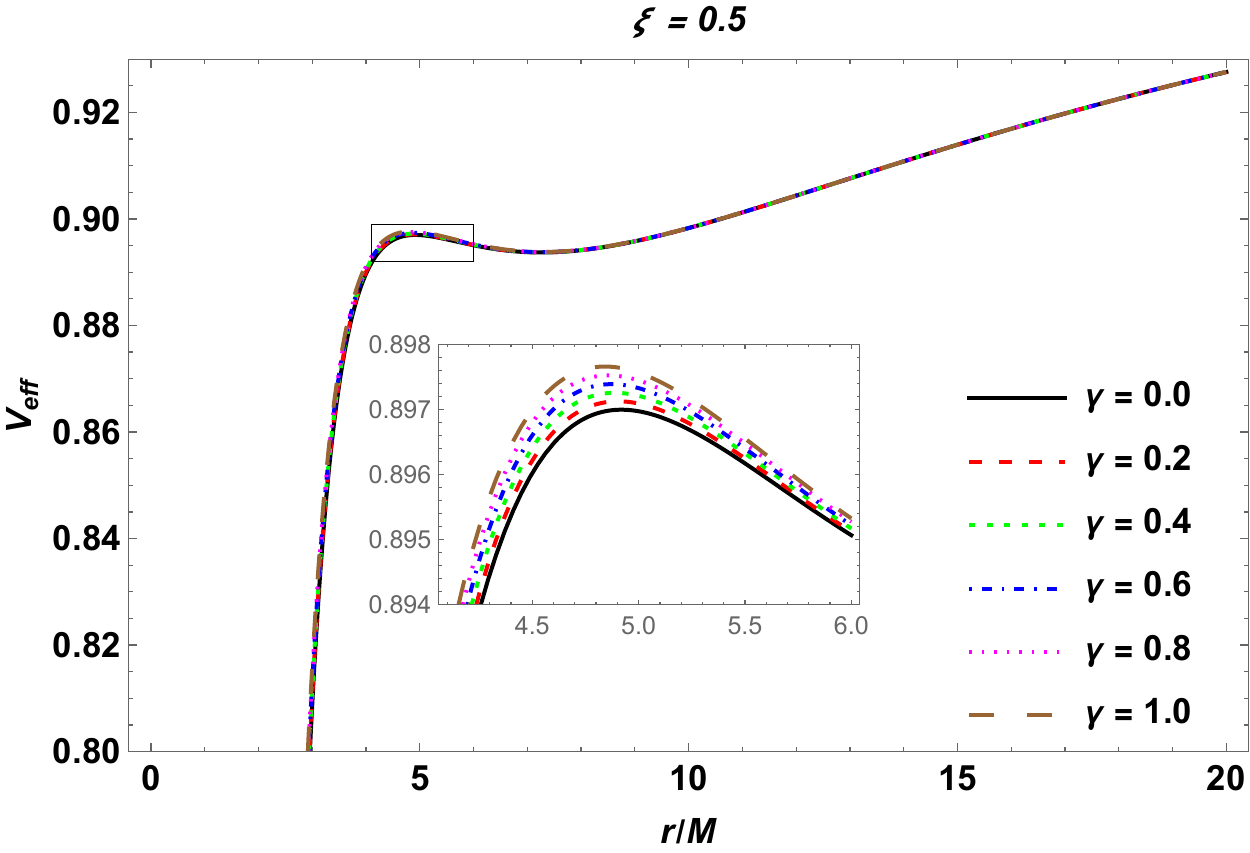}
\includegraphics[scale=0.4]{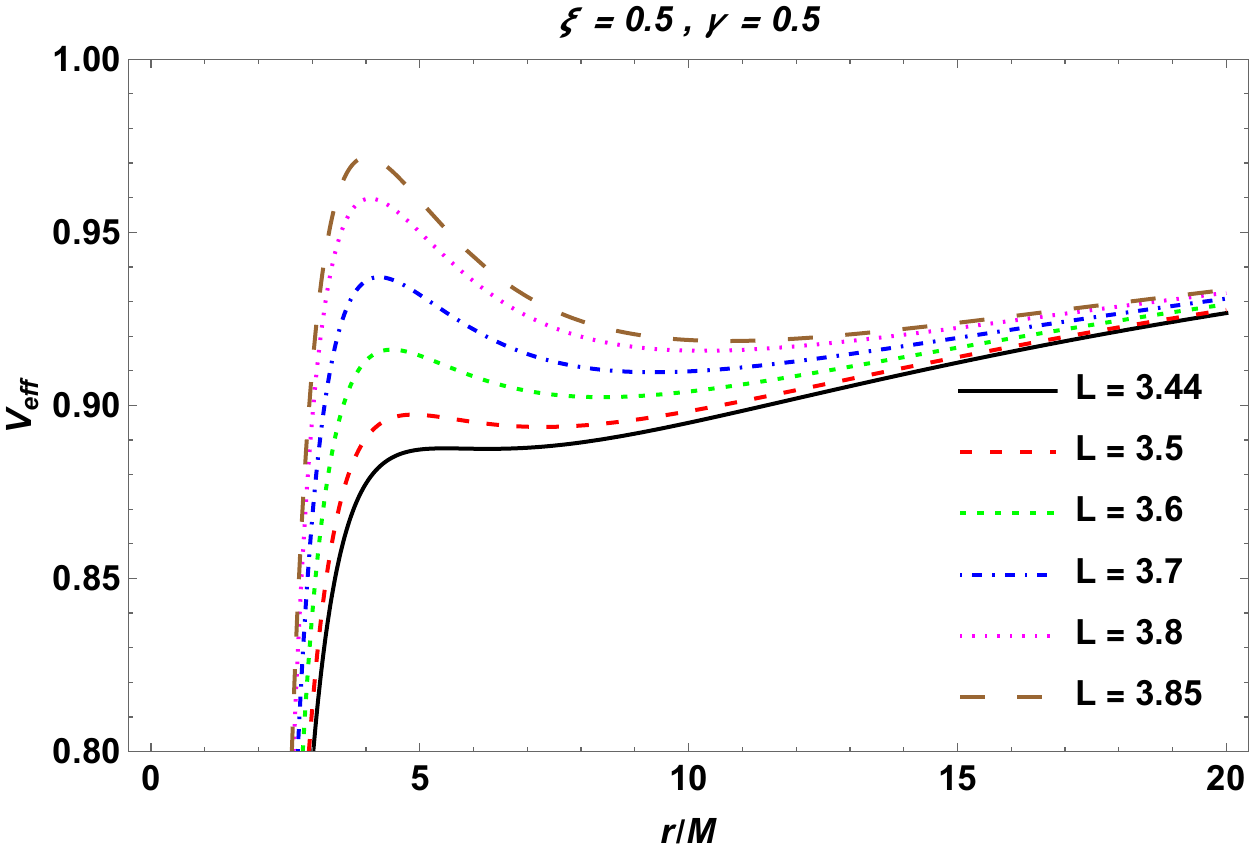}
\caption{Top panel: the radial dependence of the effective potential for the different values of the $\xi$ (left panel) and $\gamma$ (right panel) parameters. Here, we set  $\gamma=0.5$ and $\xi=0.5$ for the left and right panels, respectively. The bottom panel illustrates the radial dependence of the effective potential for the different values of the orbital angular momentum while keeping other parameters fixed as $\xi=\gamma=0.5$.}
\label{fig:eff}
\end{figure*}
\begin{figure*}
\includegraphics[scale=0.32]{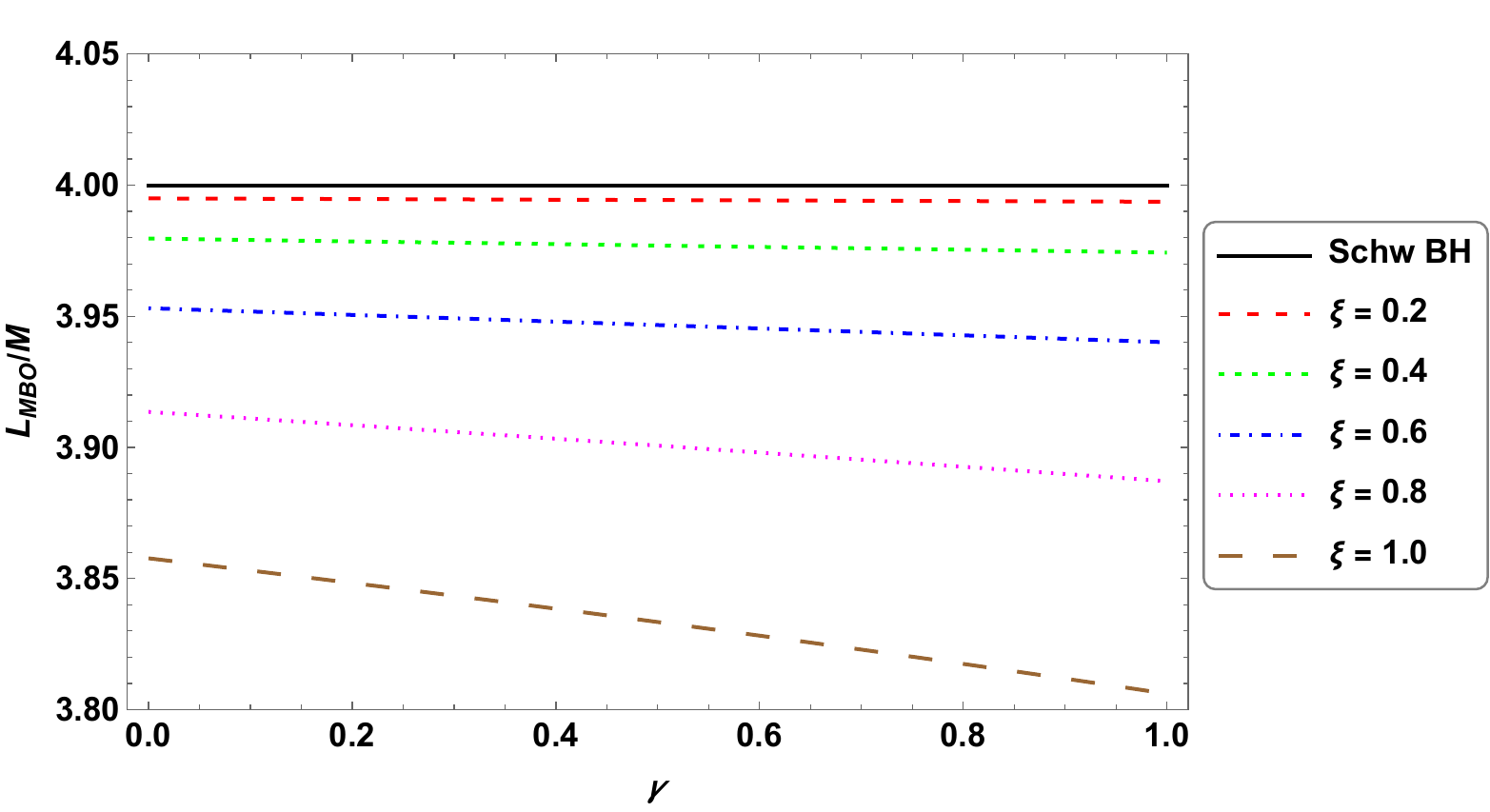}
\includegraphics[scale=0.32]{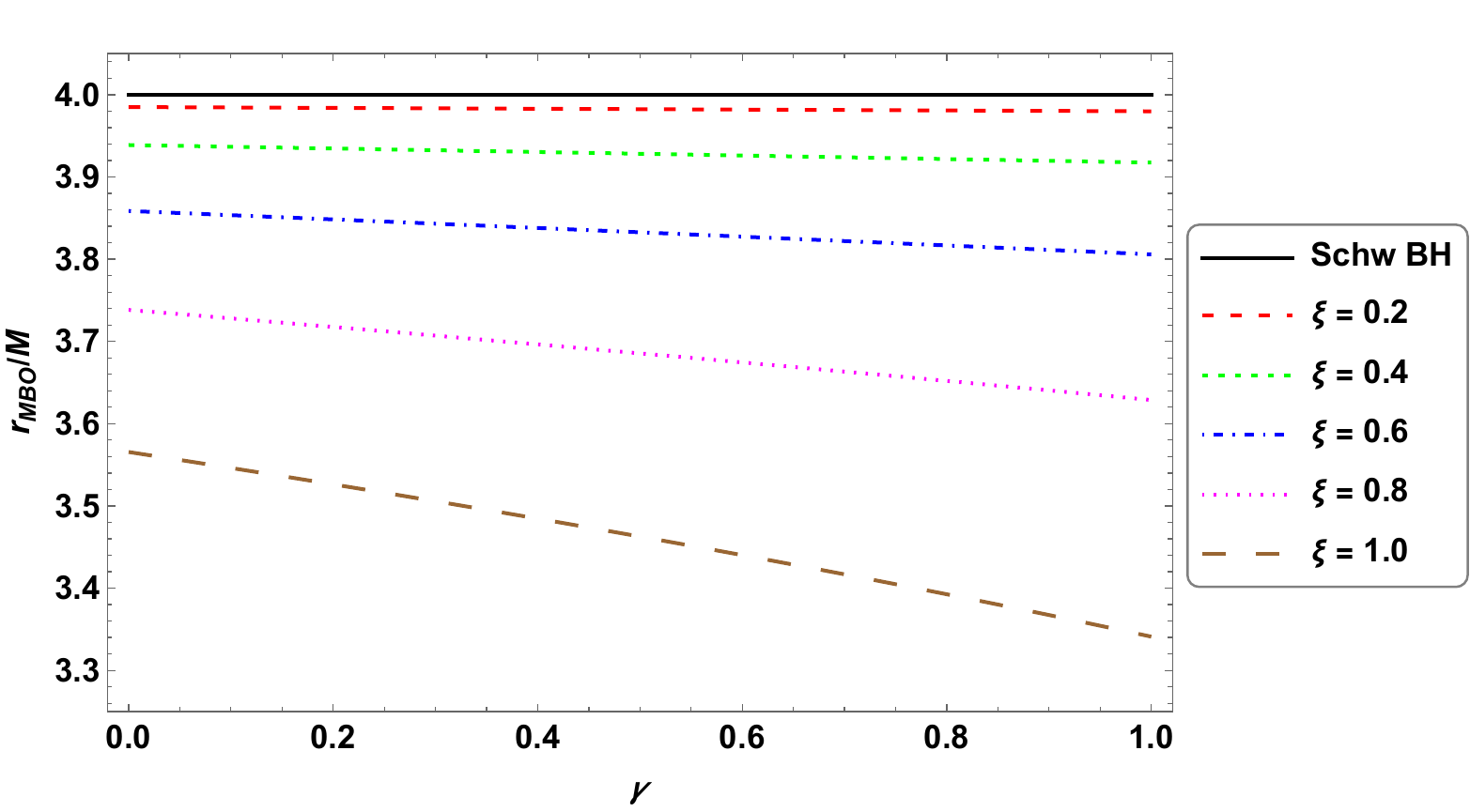}
\caption{The MBO parameters, which are orbital angular momentum and radius, as a function of the $\gamma$ parameter for the different values of the $\xi$ parameter. }
\label{fig:mbo}
\end{figure*}
\begin{figure*}
\includegraphics[scale=0.32]{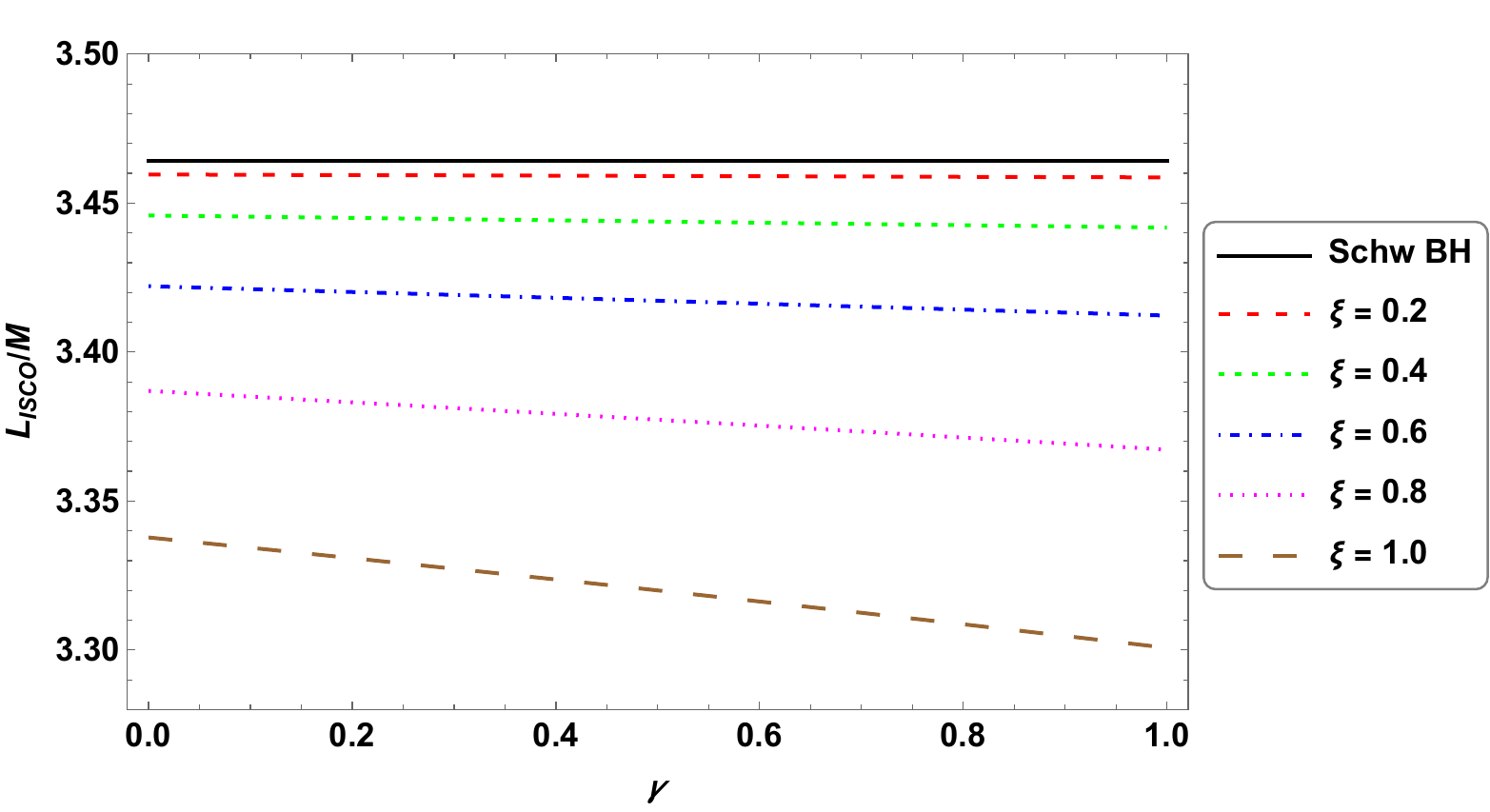}
\includegraphics[scale=0.32]{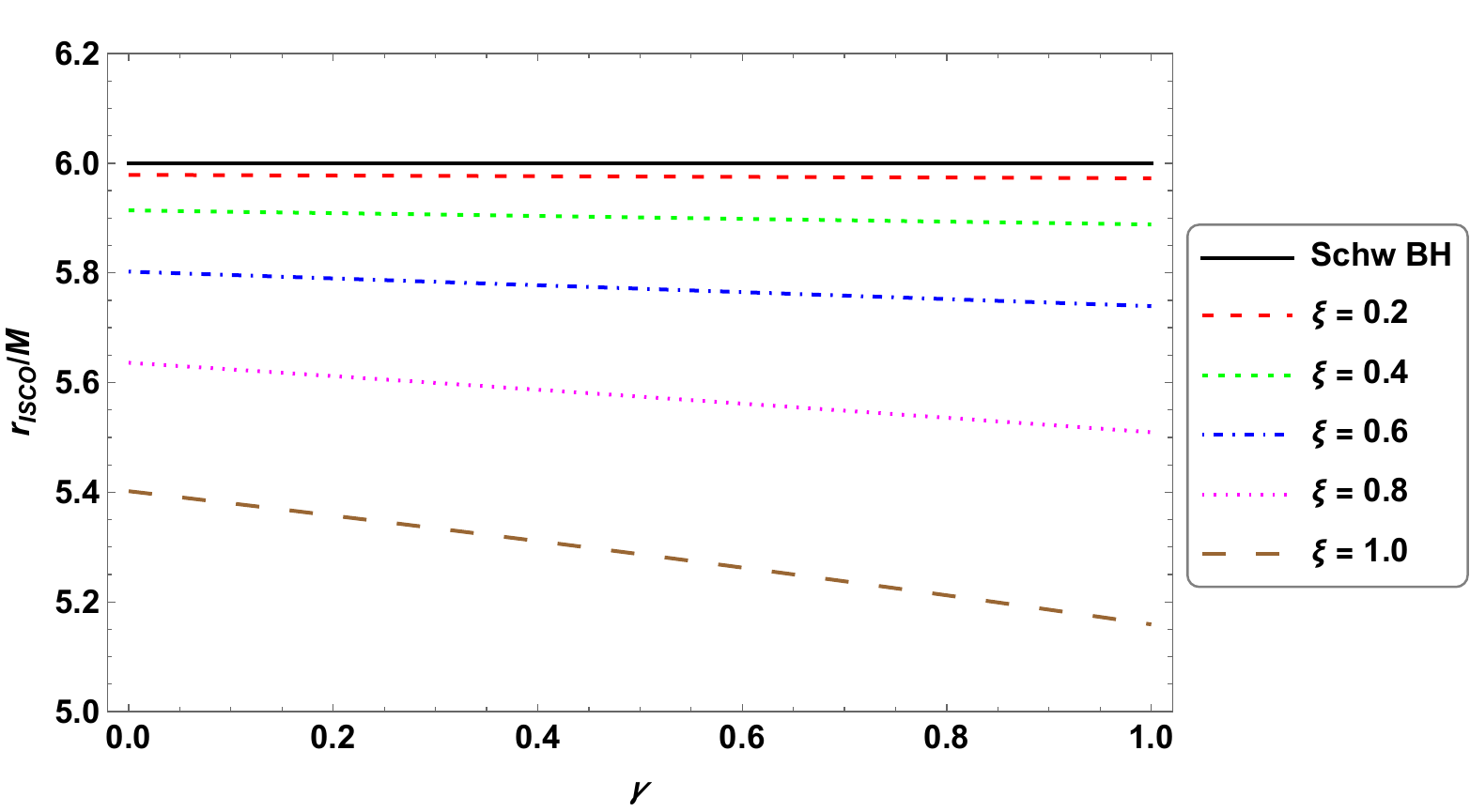}
\includegraphics[scale=0.32]{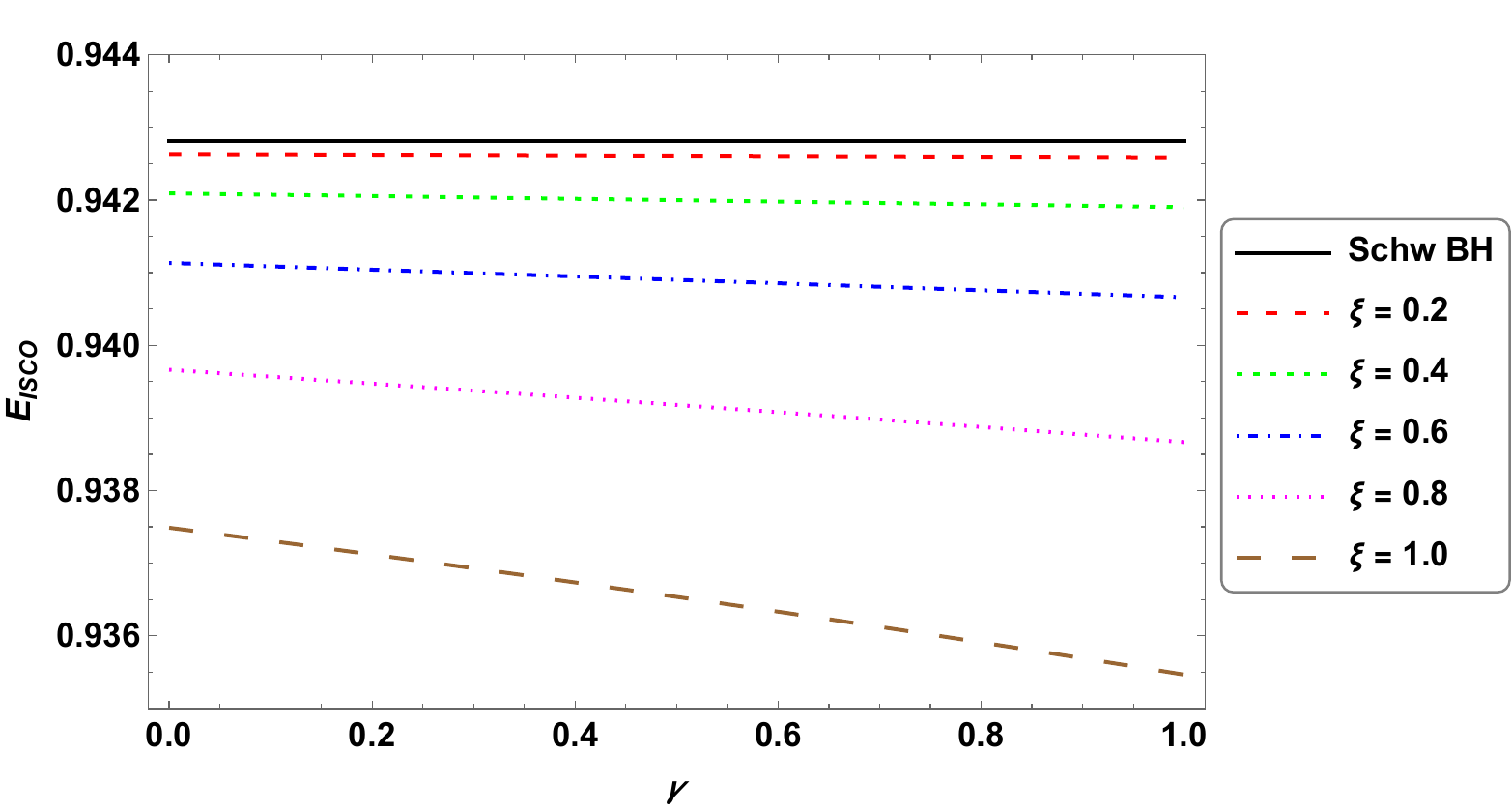}
\caption{The ISCO parameters, i.e., orbital angular momentum, radius, and energy, as a function of the $\gamma$ for different values of the $\xi$ parameter.}
\label{fig:isco}
\end{figure*}
\begin{figure*}
\includegraphics[scale=0.5]{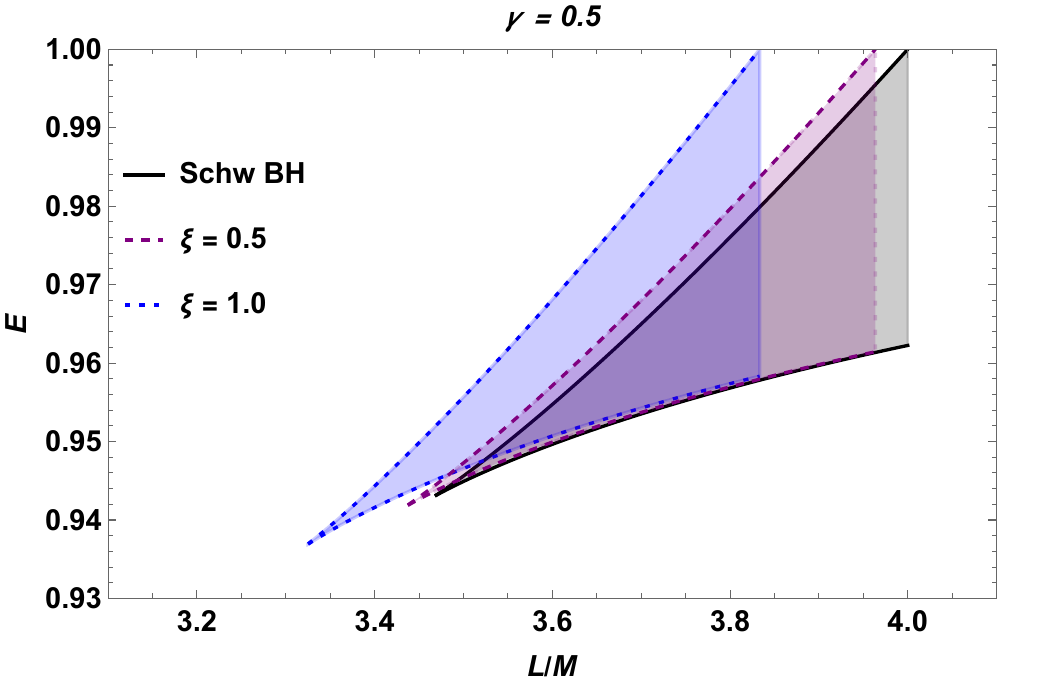}
\caption{The plot shows the allowed parameter space of the energy and orbital angular momentum for bound orbits around the new improved Schwarzschild BH with different values of the $\xi$ parameter. Here, we set $\gamma=0.5$.}
\label{fig:el}
\end{figure*}
\begin{figure*}
    \centering
    \includegraphics[width=0.32\textwidth]{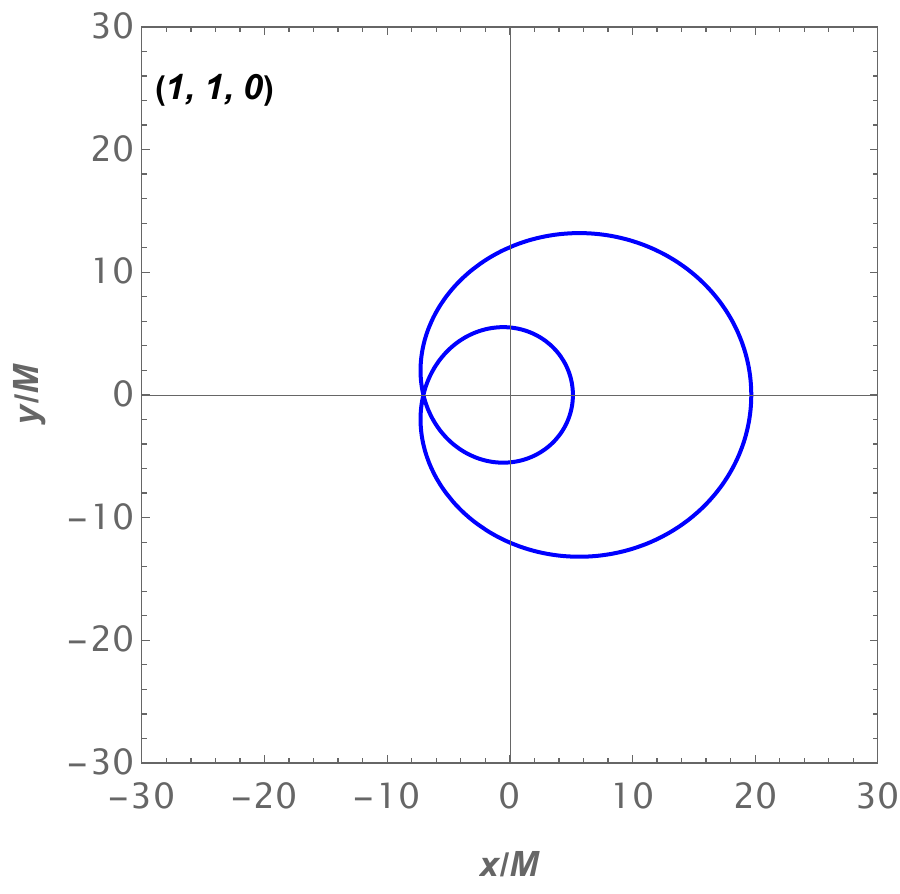} \hfill
    \includegraphics[width=0.32\textwidth]{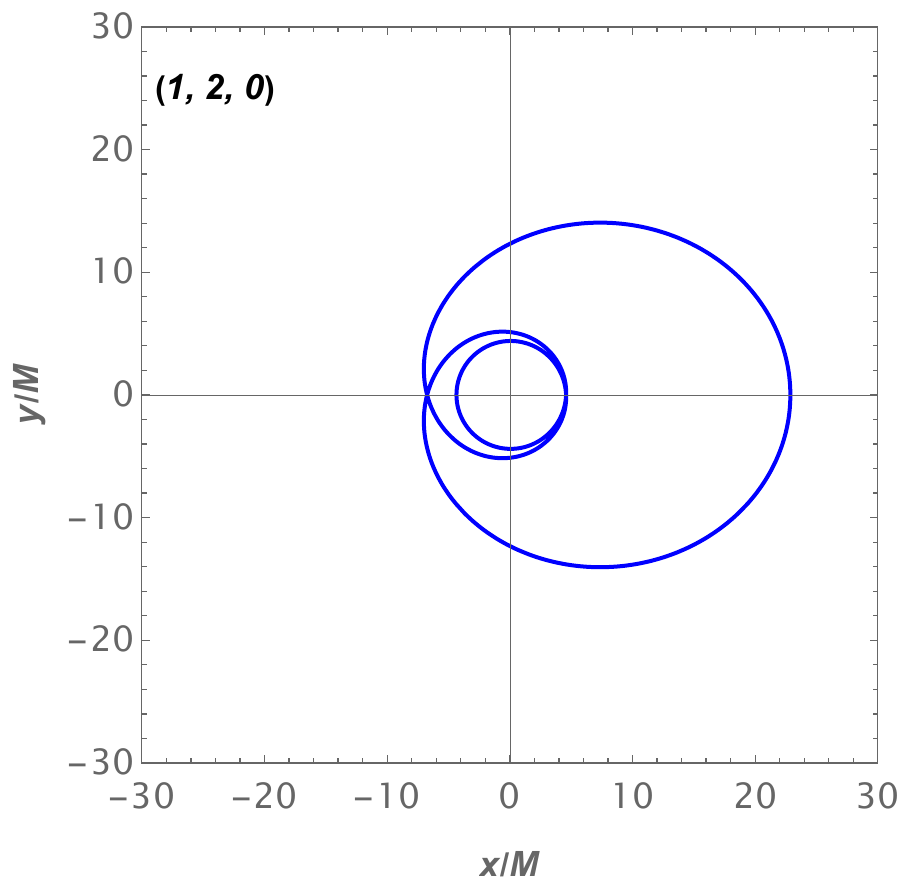} \hfill
    \includegraphics[width=0.32\textwidth]{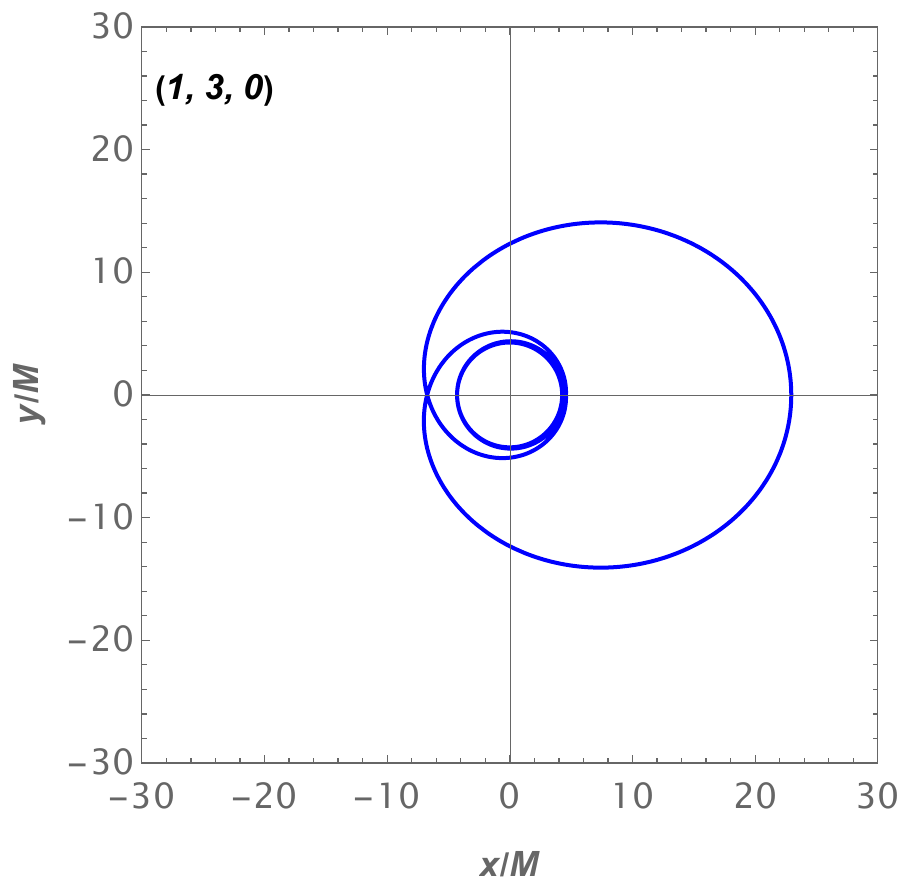} \\
    
    \vspace{0.2cm} 
    \includegraphics[width=0.32\textwidth]{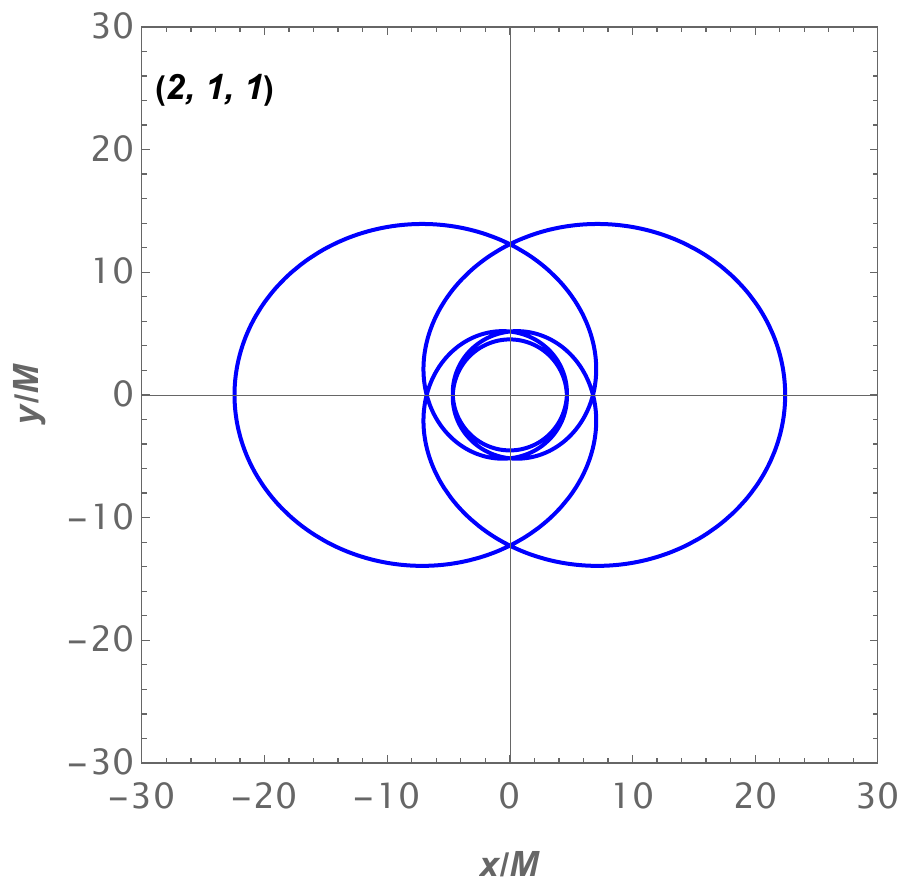} \hfill
    \includegraphics[width=0.32\textwidth]{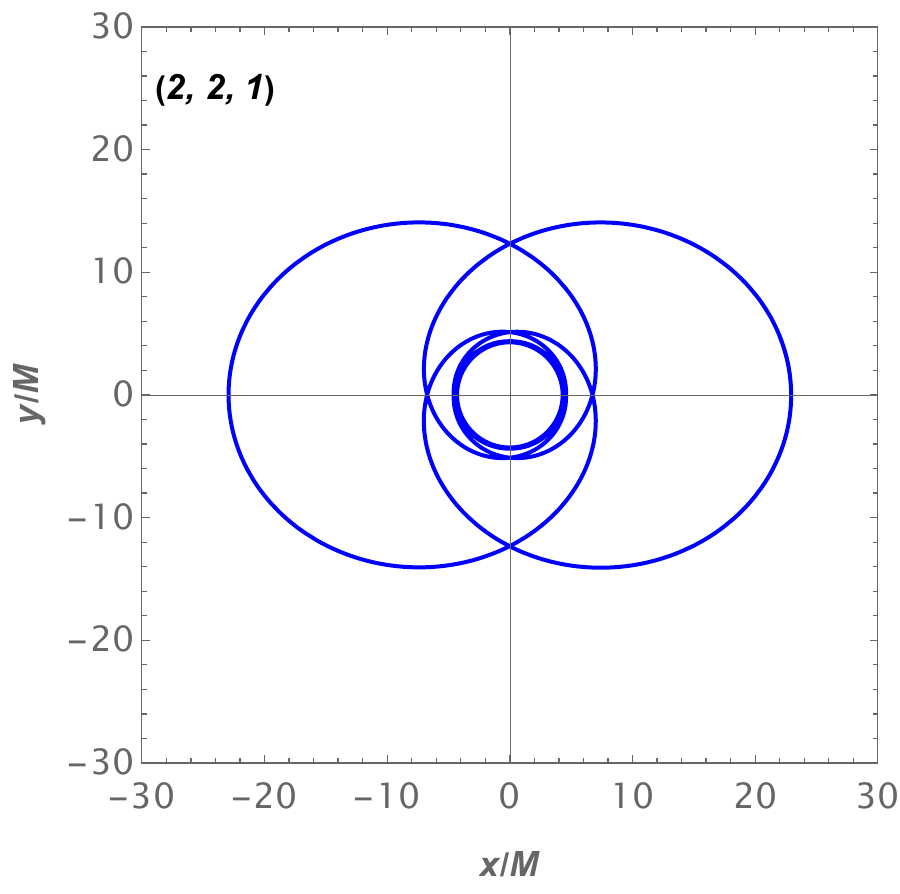} \hfill
    \includegraphics[width=0.32\textwidth]{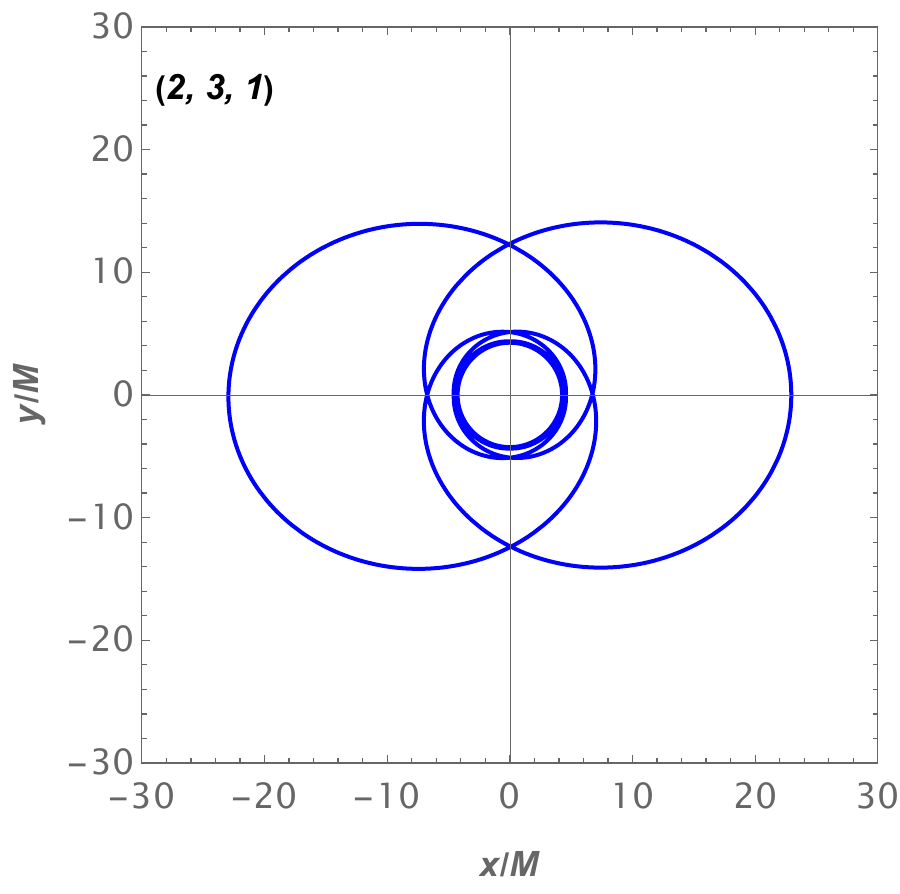} \\
    
    \vspace{0.2cm} 
    \includegraphics[width=0.32\textwidth]{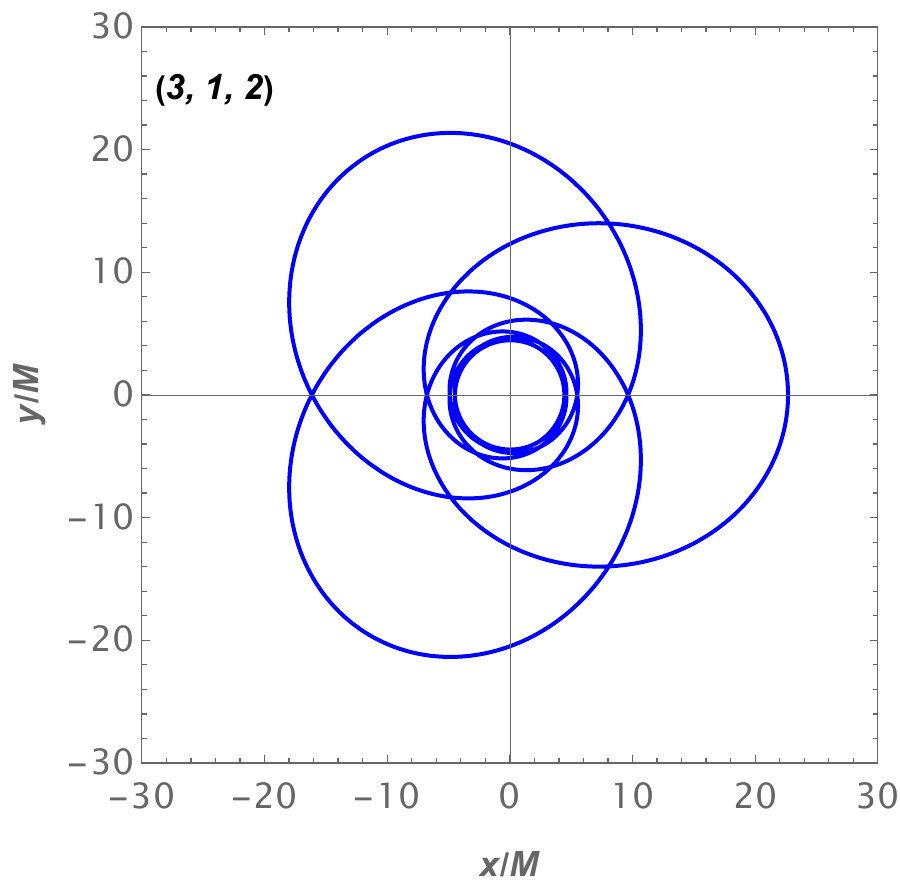} \hfill
    \includegraphics[width=0.32\textwidth]{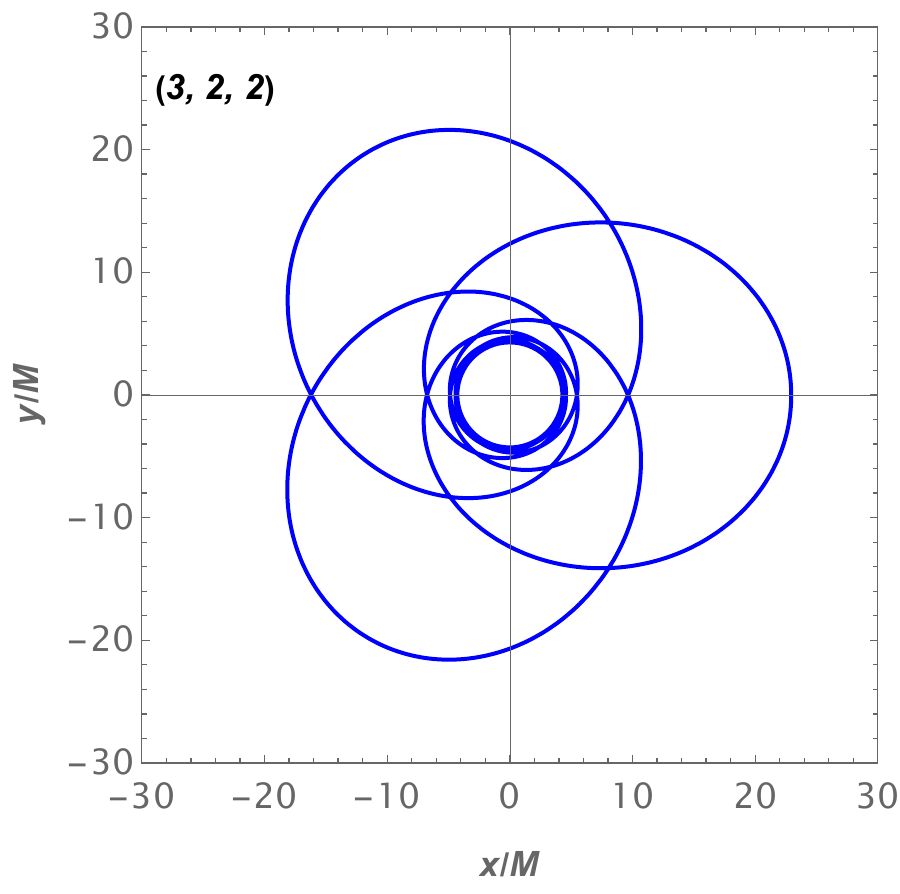} \hfill
    \includegraphics[width=0.32\textwidth]{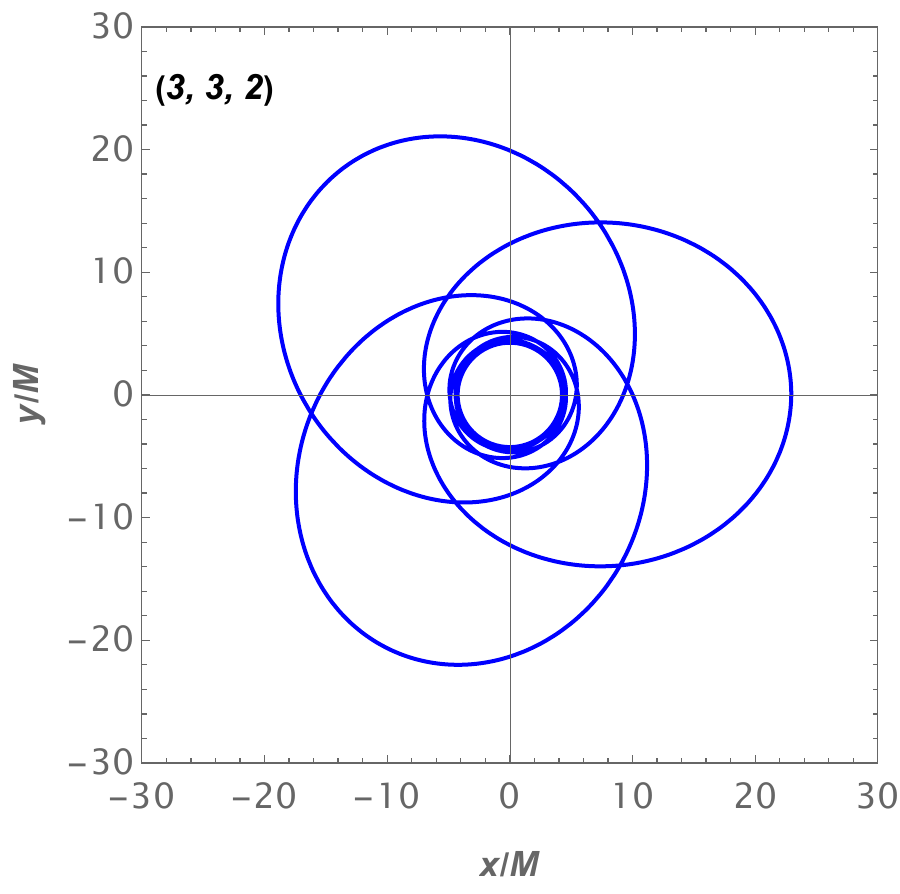} \\
    
    \vspace{0.2cm} 
    \includegraphics[width=0.32\textwidth]{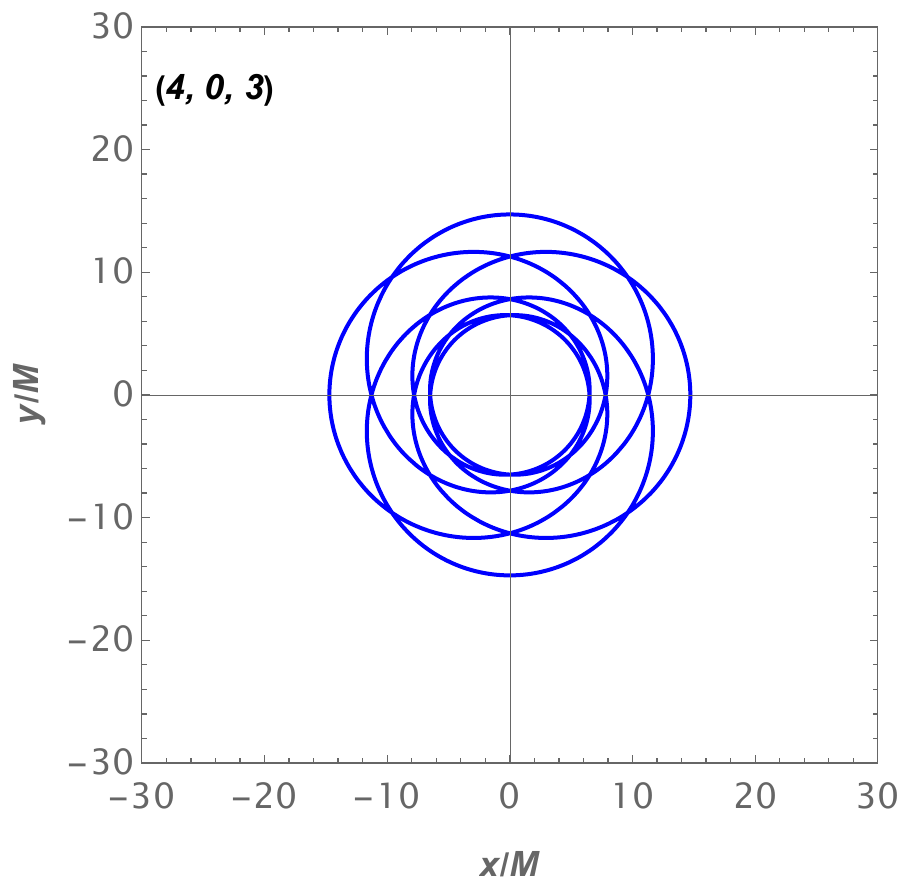} \hfill
    \includegraphics[width=0.32\textwidth]{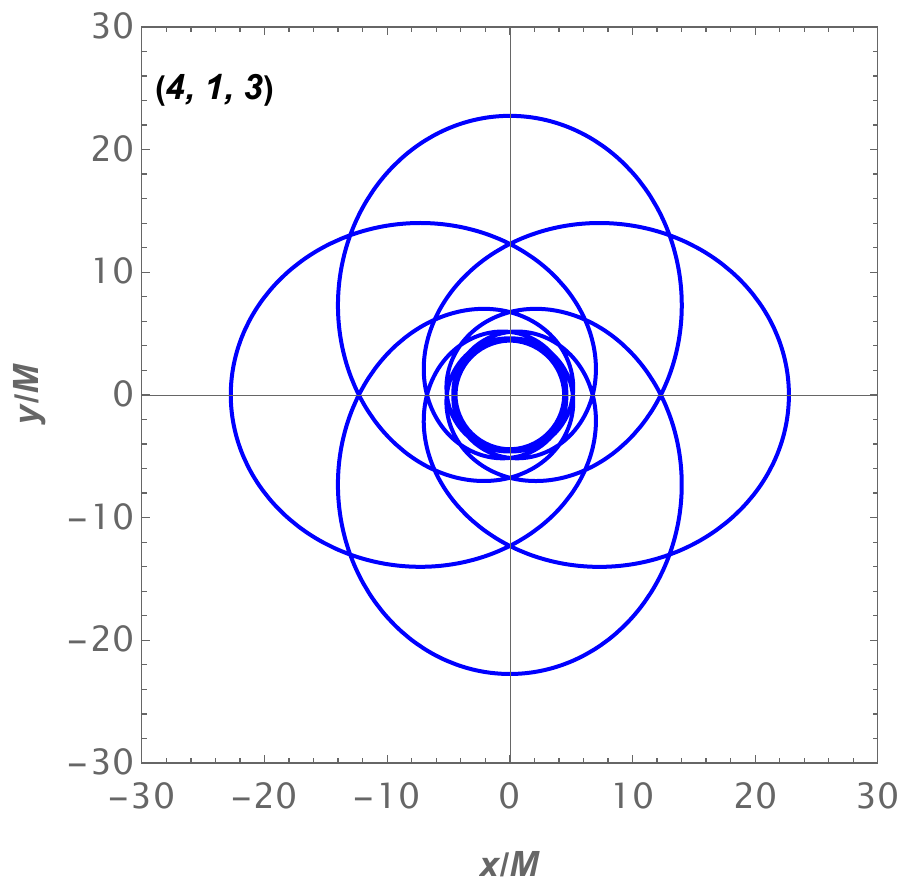} \hfill
    \includegraphics[width=0.32\textwidth]{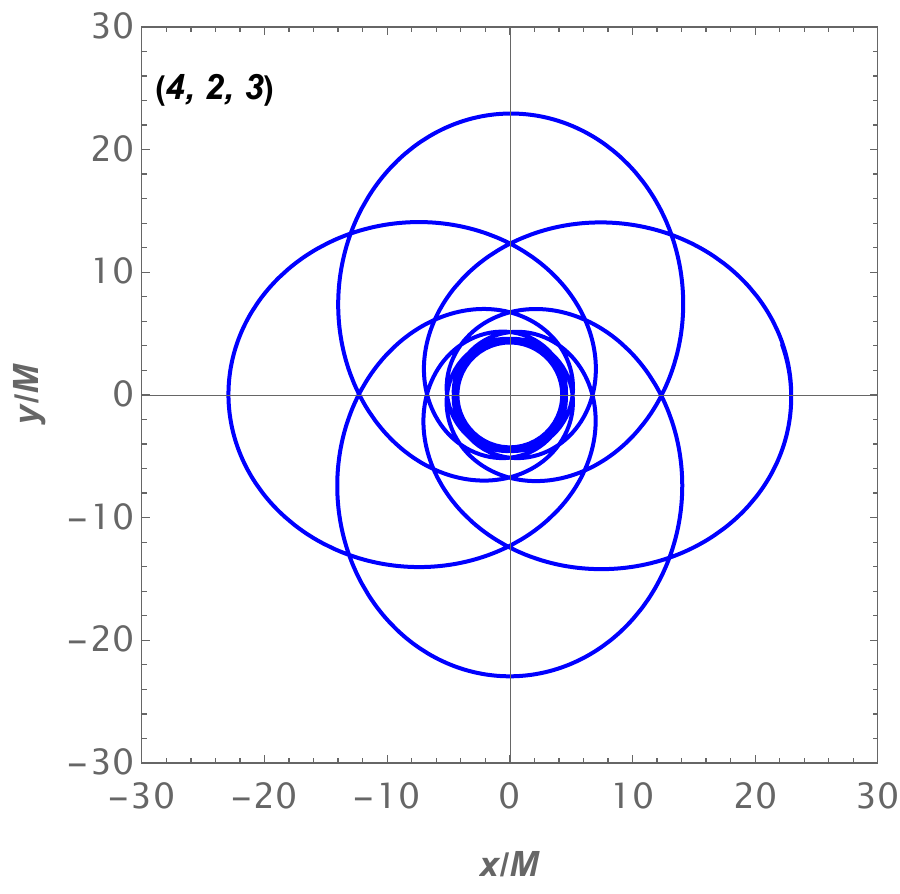}
    
    \caption{The plot shows the periodic orbits for the different combinations of zoom, whirl, and vertex numbers. We set $\xi=\gamma=0.5$ for these plots.}
    \label{fig:orbits}
\end{figure*}
\begin{figure*}[htbp]
\begin{subfigure}[b]{0.45\textwidth}
\includegraphics[width=\textwidth]{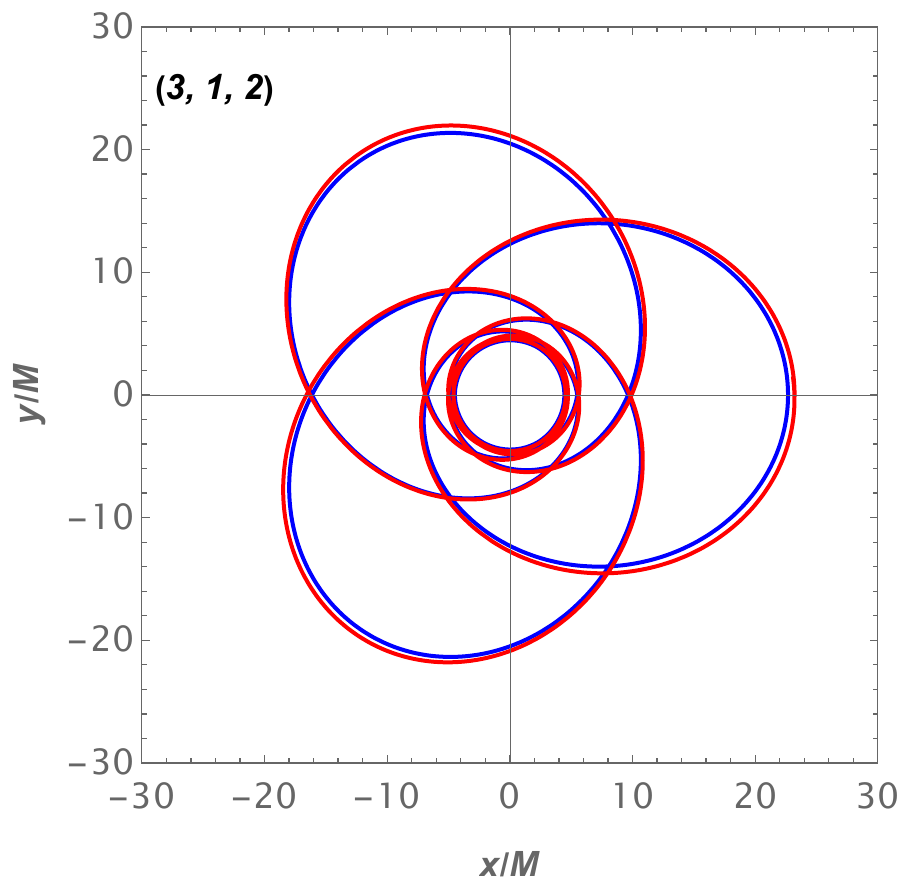}
\end{subfigure}
\begin{subfigure}[b]{0.52\textwidth}
\includegraphics[width=\textwidth]{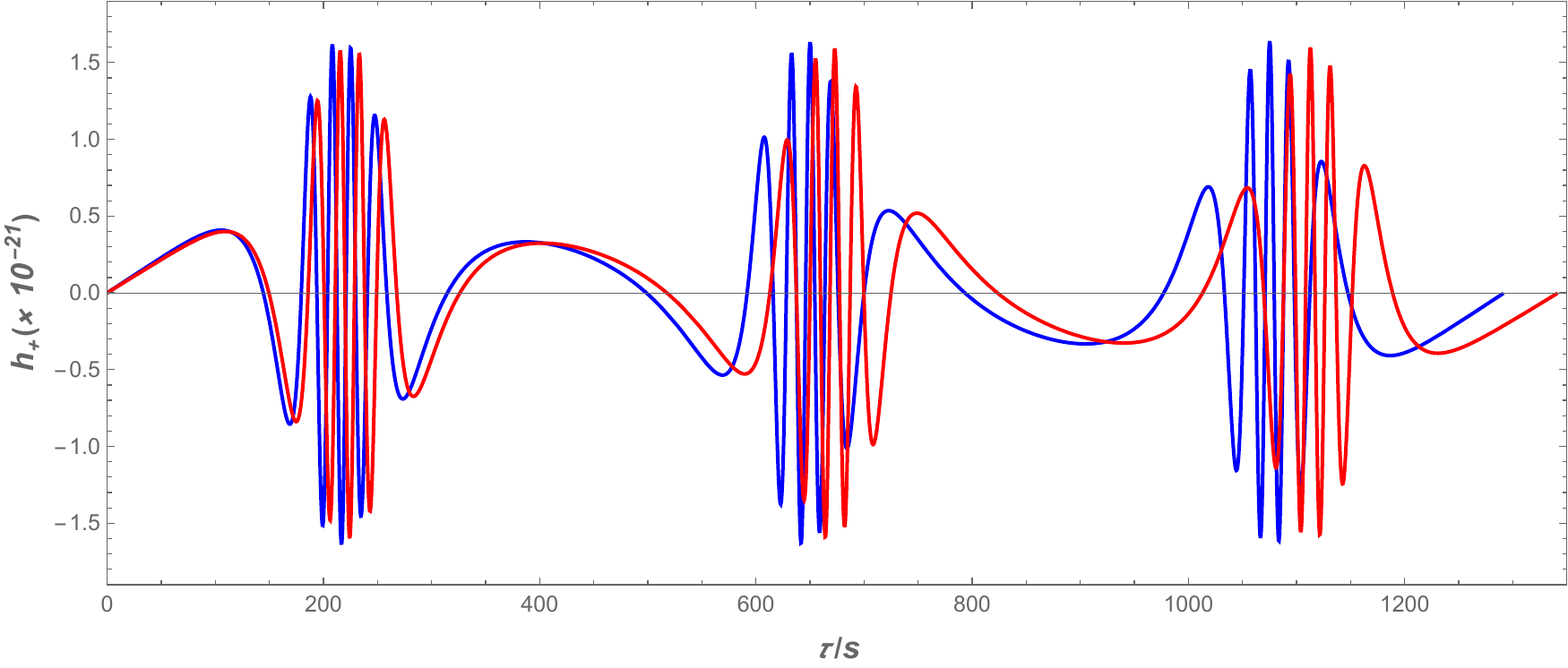}
\includegraphics[width=\textwidth]{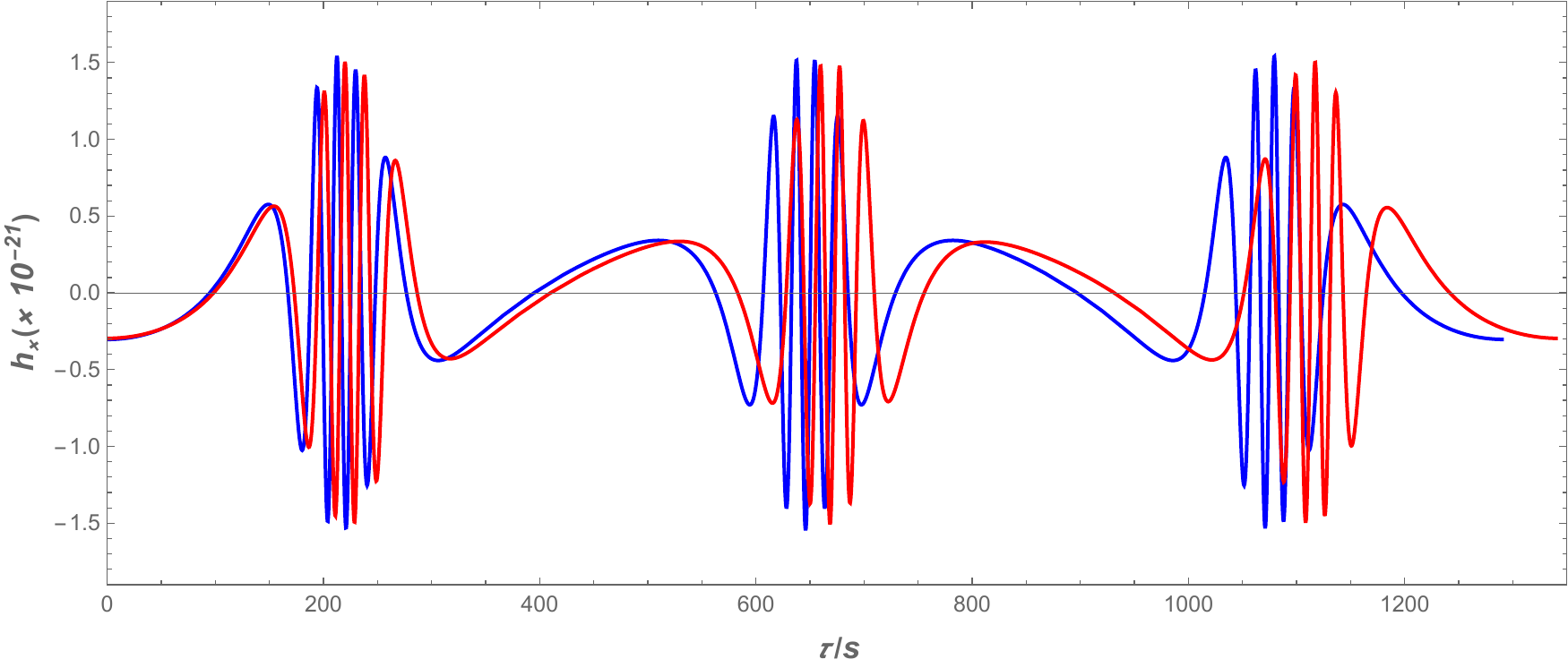}    
\end{subfigure} 
\caption{The plot represents the $(3,1,2)$ periodic orbit and associated gravitational waveforms from the EMRI system. The red and blue lines, respectively, correspond to the results for the standard Schwarzschild BH and the new improved Schwarzschild BH with $\xi=\gamma=0.5$.}
\label{fig:comb}
\end{figure*}
\begin{figure*}
\begin{subfigure}[b]{0.45\textwidth}
\includegraphics[width=\textwidth]{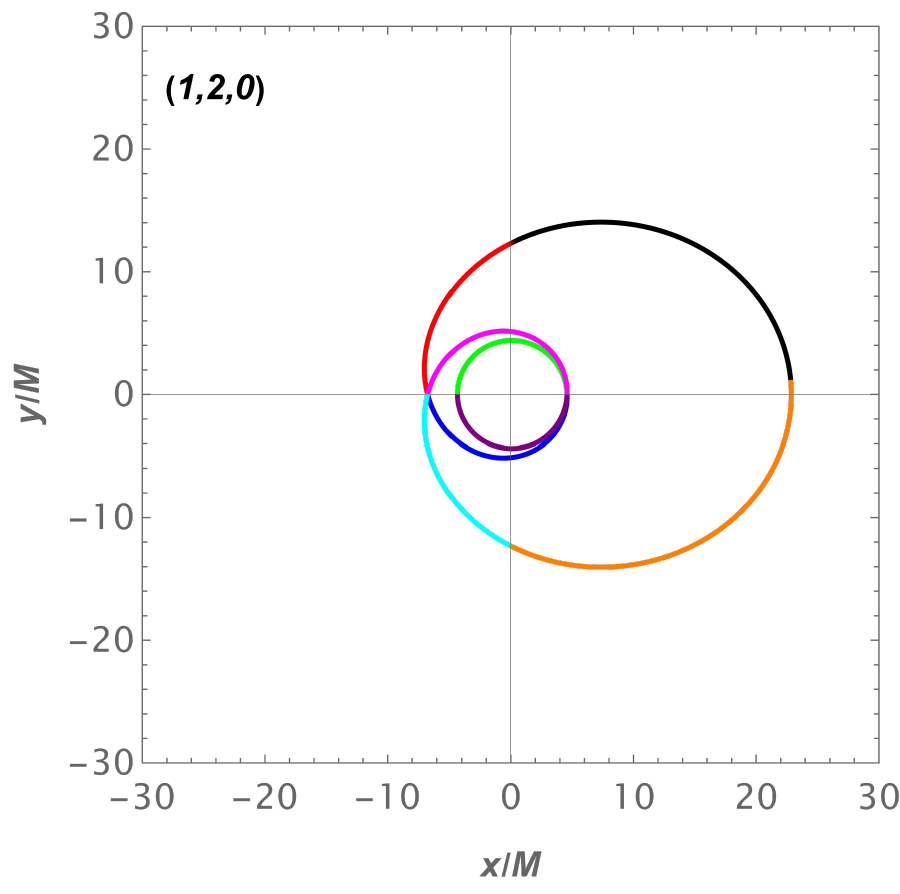}
\end{subfigure}
\begin{subfigure}[b]{0.52\textwidth}
\includegraphics[width=\textwidth]{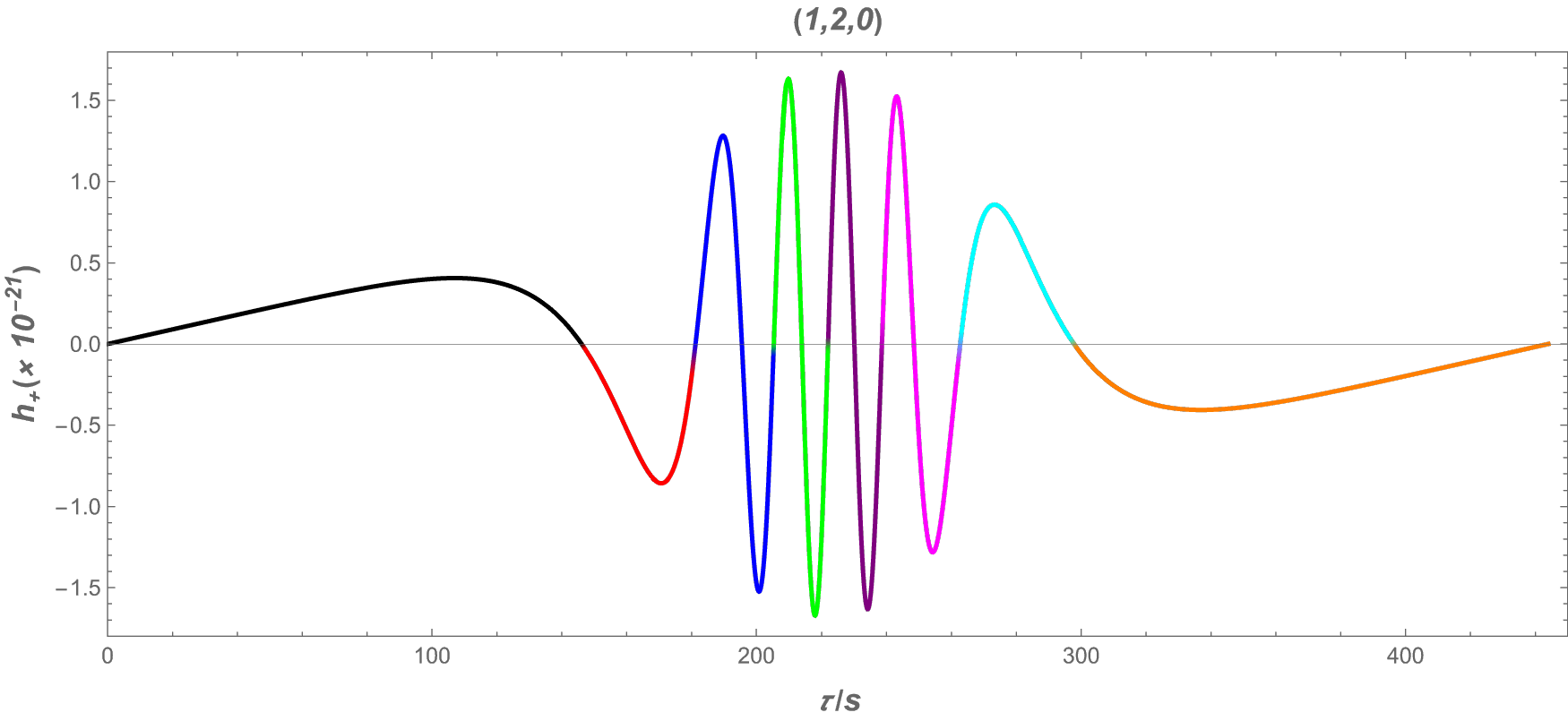}
\includegraphics[width=\textwidth]{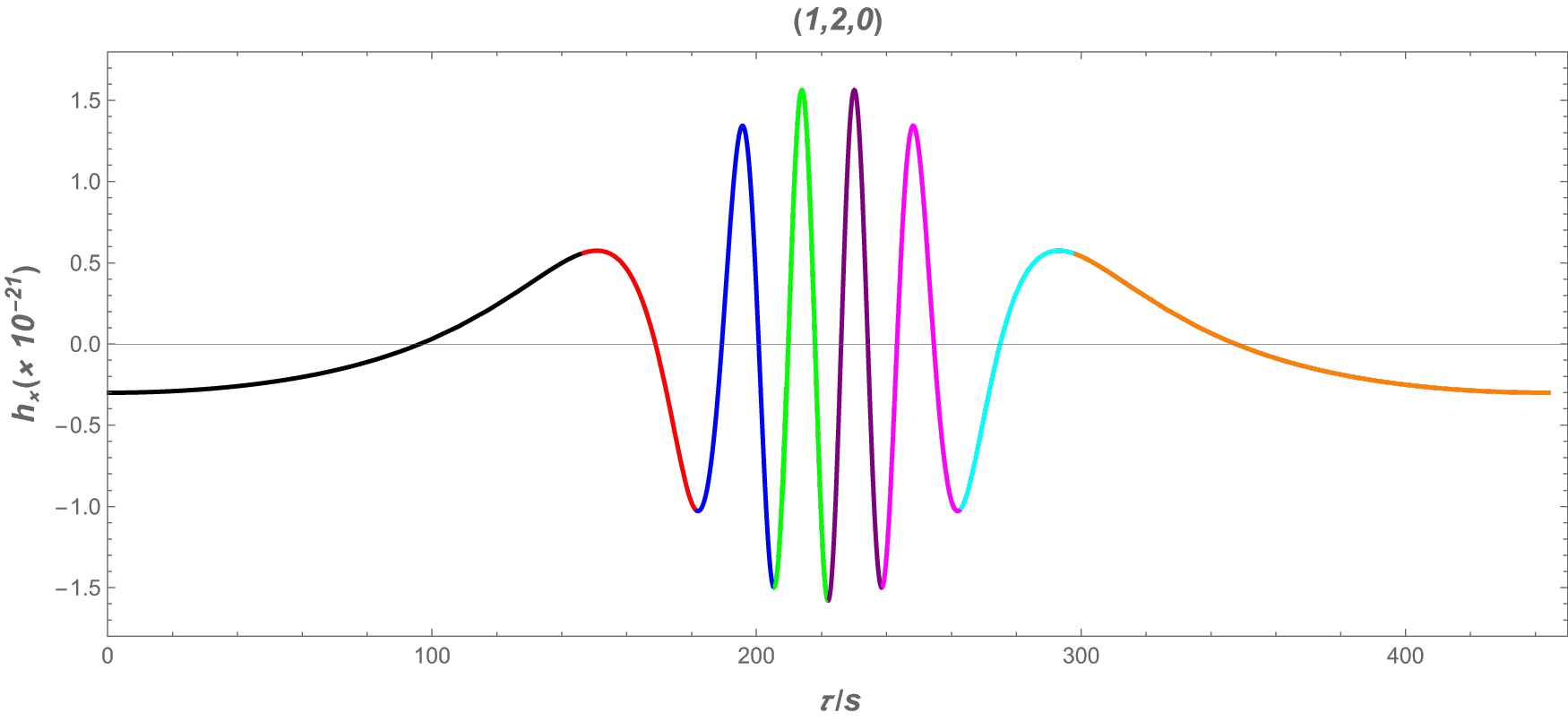}    
\end{subfigure} 
\caption{The figure depicts the ($1,2,0$) periodic orbit and its associated gravitational waveforms, with each segment distinguished using different colors. Here, we set $\xi=\gamma=0.5$.}
\label{fig:120}
\end{figure*}
\begin{figure*}
\includegraphics[scale=0.5]{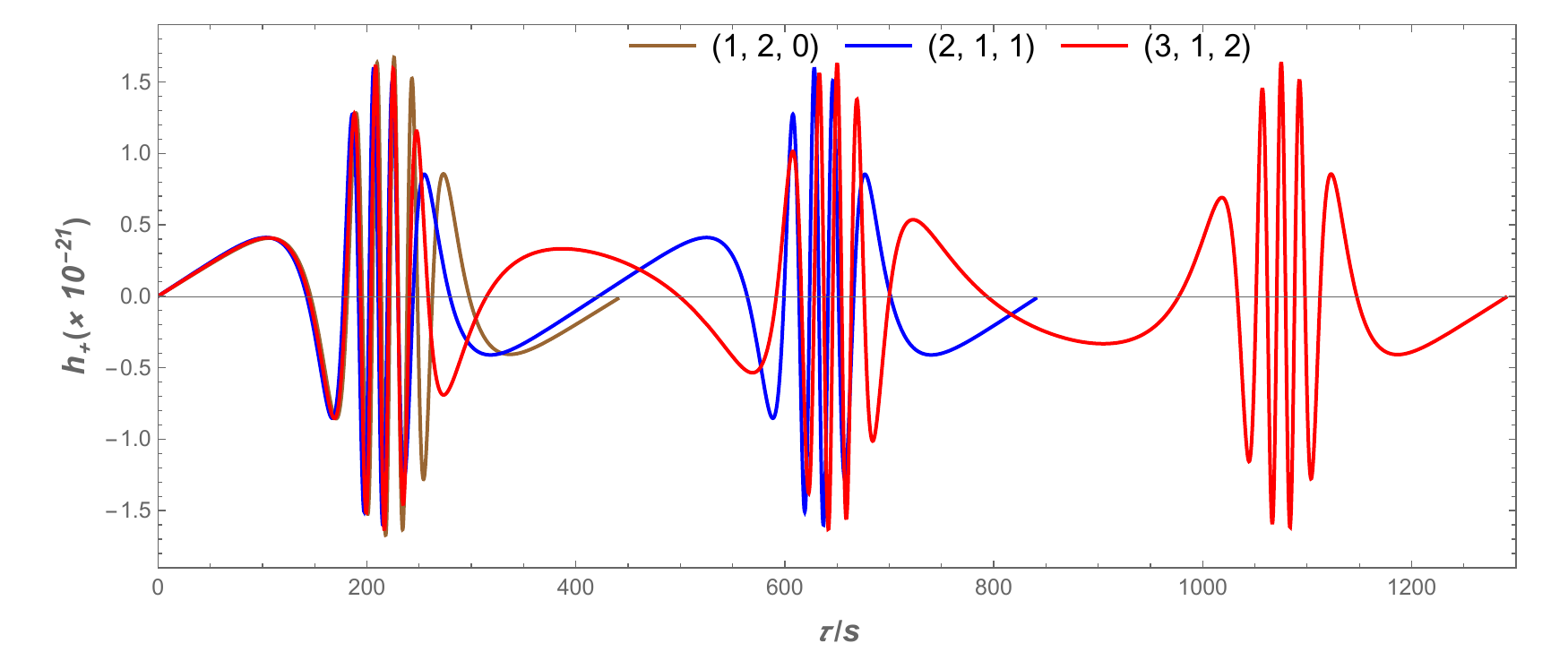}
\includegraphics[scale=0.5]{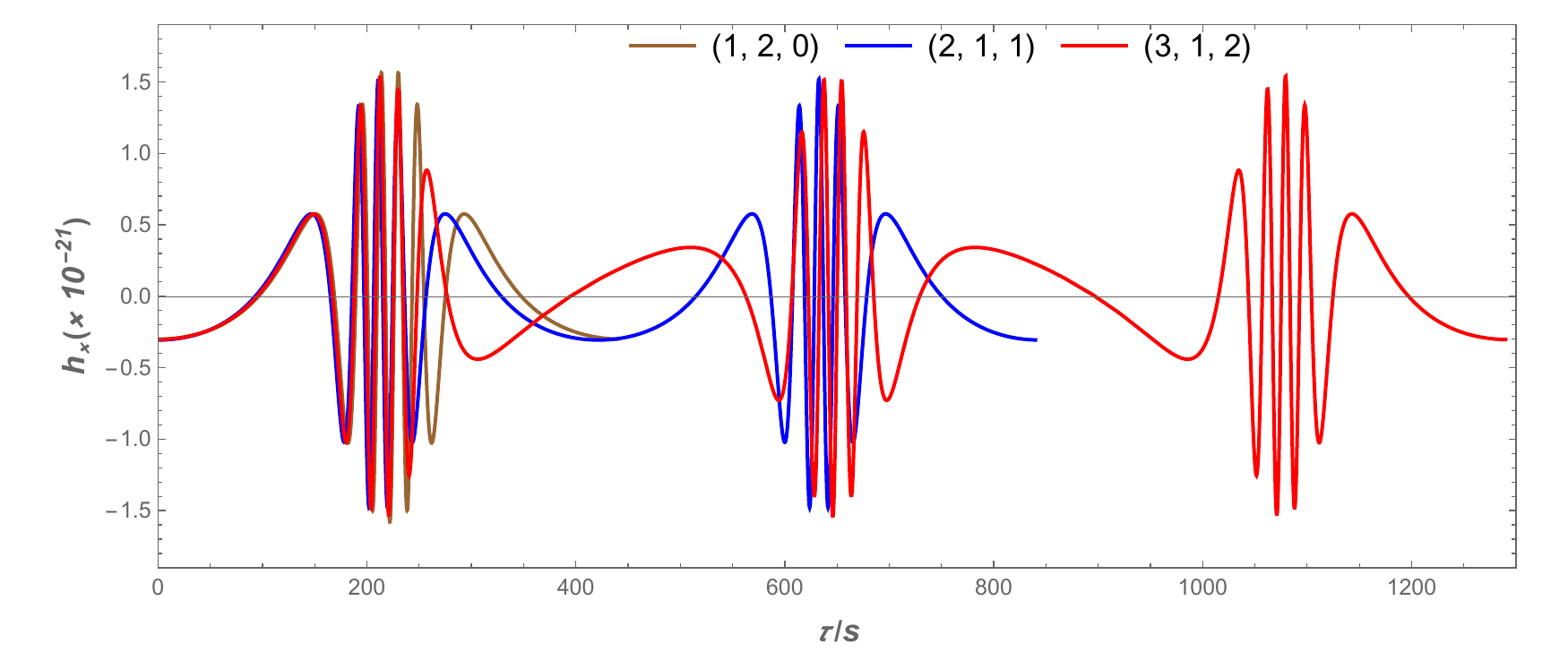}
\caption{The figure compares the gravitational waveforms generated by different periodic orbits. The spacetime parameters were set as $\xi=\gamma=0.5$.}
\label{fig:comp}
\end{figure*}

This section is devoted to a brief review of the spacetime of the new improved Schwarzschild BH and massive particle dynamics around it. One can write the line element as follows~\cite{ALENCAR2026170756}
\begin{equation}
ds^2=-f(r)dt^2+\frac{dr^2}{f(r)}+r^2d\theta^2+r^2\sin^2{\theta}d\phi^2\,,
\end{equation}
where
\begin{equation}\label{metric}
    f(r)=1-\frac{4Mr^2}{\xi^2(\gamma M+r)+\sqrt{\xi^4(\gamma M+r)^2+4r^6}}\,.
\end{equation}
The metric functions contain additional $\xi$ and $\gamma$ parameters besides the BH mass. These parameters, respectively, denote the cutoff scale and interpolation parameters. It can be seen from Eq.~(\ref{metric}) that one can recover the Schwarzschild spacetime when the $\xi$ parameter equals zero. Therefore, we compare our results with those of the Schwarzschild BH in subsequent parts. To analyze the behaviour of the metric function $f(r)$, we plot its radial profile for different values of $\xi$ and $\gamma$ in Fig.~\ref{fig:metric}. Keeping in mind that $f(r)=0$ defines the horizons of BH, we can state that the Cauchy and event horizons increase under the influence of the parameters $\xi$ and $\gamma$. 

Now, we can turn to the investigation of test particle dynamics around the new improved Schwarzschild BH by utilizing the Lagrangian formalism. One can write the Lagrangian of the test particle with mass $m$ in the following way~\cite{1983mtbh.book.....C}
\begin{equation}
\mathcal{L}=\frac{1}{2}mg_{\mu\nu}\frac{dx^{\mu}}{d\tau}\frac{dx^{\nu}}{d\tau}\,,
\end{equation}
where $\tau$ denotes the proper time. From the above equation, we can find the generalized momentum by setting $m=1$ for simplicity, as
\begin{equation}
p_{\mu}=\frac{\partial \mathcal{L}}{\partial \dot{x}^{\mu}}=g_{\mu\nu}\dot{x}^{\nu}\,.
\end{equation}
The above equation takes the following form when we consider constants of motion
\begin{eqnarray}
p_t&=&-\Big(1-\frac{4Mr^2}{\xi^2(\gamma M+r)+\sqrt{\xi^4(\gamma M+r)^2+4r^6}}\Big) \dot{t}=-E\,, \nonumber\\
p_{\phi}&=&r^2\sin^2{\theta}\dot{\phi}=L\,, \nonumber \\
p_{r}&=&\Big(1-\frac{4Mr^2}{\xi^2(\gamma M+r)+\sqrt{\xi^4(\gamma M+r)^2+4r^6}}\Big)^{-1}\dot{r}\,,\nonumber \\
p_{\theta}&=&r^2\dot{\theta}\,,
\end{eqnarray}
where $L$ and $E$ refer to the orbital angular momentum and energy, respectively. By utilizing the normalization condition and equatorial plane motion, i.e., $g_{\mu\nu}\dot{x}^{\mu}\dot{x}^{\nu}=-1$ and $\theta=\pi/2$, one can find the following equation 
\begin{equation}
    \dot{r}^2+V_{eff}=E^2\,.
\end{equation}
In the above equation, $V_{eff}$ serves as the effective potential. It can be written explicitly as
\begin{eqnarray}
    V_{eff}&=&\left(1-\frac{4Mr^2}{\xi^2(\gamma M+r)+\sqrt{\xi^4(\gamma M+r)^2+4r^6}}\right)\times \nonumber \\
    &\times& \left(1+\frac{L^2}{r^2}\right)\,.
\end{eqnarray}
In Fig.~\ref{fig:eff}, we plot the radial profile of the effective potential for the different values of $\xi$, $\gamma$, and orbital angular momentum $L$. As shown in this figure, the values of the effective potential increase under the effect of the $\xi$ parameter and $L$. Also, there is a slight increase with the increase of the $\gamma$ parameter.
According to the values of the energy and angular momentum of particles around the BH, there are different types of orbits. However, in this work, we focus on the periodic orbits. Therefore, we must consider the following conditions for the periodic orbits~\cite{Dadhich22a}:
\begin{align}
L_{ISCO}\leq L \quad \text{and} \quad E_{ISCO}\leq E \leq E_{MBO}=1\,.
\end{align}
In the above equation, $L_{ISCO}$ and $E_{ISCO}$ denote the angular momentum and energy of the particle orbiting on the ISCO, while $E_{MBO}$ is the energy of the particle orbiting along the MBO. Hence, in order to study periodic orbits, it is important to explore the MBO and ISCO. The MBO can be determined by the following conditions~\cite{2018igrc.book.....B}:
\begin{align}
    V_{eff}=1\,, \quad \frac{dV_{eff}}{dr}=0\,.
\end{align}
Due to the complex form of the effective potential, it is hard to obtain the MBO parameters analytically. We hence choose the numerical approach. In Fig.~\ref{fig:mbo}, we demonstrate the $L_{MBO}$ and $r_{MBO}$ as a function of $\gamma$ for different values of the $\xi$ parameter. Both MBO parameters decrease with an increase in the spacetime parameters, which are $\xi$ and $\gamma$. The black line represents the results for the Schwarzschild BH. We now turn to the ISCO by considering the following conditions~\cite{Hartle_2021}
\begin{align}
\dot{r}=0\,, \quad \frac{dV_{eff}}{dr}=0 \quad \text{and} \quad \frac{d^2V_{eff}}{dr^2}=0\,.
\end{align}
By utilizing the numerical approach, the ISCO parameters were plotted as functions of the $\gamma$ parameter for the various values of the $\xi$ parameter. It is evident that all ISCO parameters diminish in response to the effects of the spacetime parameters. 

Using the corresponding bound-orbit conditions, one can plot the allowed parameter space of energy and orbital angular momentum for the different values of the spacetime parameters as demonstrated in Fig.~\ref{fig:el}. It can be seen from this figure that the region shifts toward smaller values of the orbital angular momentum as the $\xi$ parameter increases. In other words, for a fixed orbital angular momentum, bound orbits around the new improved Schwarzschild BH can have higher energies for larger values of $\xi$. It should be noted that varying the $\gamma$ parameter may lead to a very small change in this region.

\section{Periodic orbits around the new improved Schwarzschild black hole}\label{sec3}

In this part, we study the periodic orbits of massive particles around the new improved Schwarzschild BH. It is important to note that each periodic orbit is defined by the zoom, whirl, and vertex numbers. Therefore, one can define the rational number, ratio of the angular $\omega_{\phi}$ and radial $\omega_r$ frequencies, as~\cite{Levin_2008}
\begin{eqnarray}
 q=\frac{\omega_{\phi}}{\omega_r}-1=w+\frac{v}{z}\,,
\end{eqnarray}
where $z$, $w$ and $v$ are integers, where $v<z$ condition is satisfied for them. If we consider the equations of motion, the above equation can be read as~\cite{Shabbir:2025kqh, Jiang:2024cpe, Alloqulov:2025ucf}
\begin{eqnarray}
q&=&-1+\int_{r_1}^{r_2}\frac{\dot{\phi}}{\dot{r}}dr=\nonumber \\
&=&-1+\frac{1}{\pi}\int_{r_1}^{r_2}\frac{1}{r\sqrt{E^2-\Big(1-\frac{4Mr^2}{\xi^2(\gamma M+r)+\sqrt{\xi^4(\gamma M+r)^2+4r^6}}\Big)}}dr\,,
\end{eqnarray}
with $r_1$ and $r_2$ are, respectively, the periapsis and apoapsis radii of the periodic orbits. Now, one can demonstrate the different periodic orbits by setting the orbital angular momentum as $L=\frac{1}{2}(L_{MBO}+L_{ISCO})$. It should be emphasized that here, we use the bisection method to solve complex equations in order to define the energy associated with different periodic orbits. Using the numerical values, we plot the periodic orbits for the different values of $z$, $w$ and $v$ in Fig.~\ref{fig:orbits}. This figure reveals that massive particles with higher zoom number $z$ trace periodic orbits exhibiting progressively more complex structural patterns, whereas particles with larger whirl number $w$ execute additional loops around the BH before reaching the subsequent apoapsis.

\section{Gravitational waveforms}\label{sec4}

In this part, we study the gravitational waveforms from the periodic orbits around the new improved Schwarzschild BH. To do so, we take into account the EMRI system consisting of a stellar-mass object orbiting around the supermassive new improved Schwarzschild BH. By analyzing the gravitational waveforms produced by this system, one can define the signature of the spacetime. It is possible to evaluate the gravitational waveforms by considering the adiabatic approximation, a widely adopted method. On short timescales, energy and angular momentum remain nearly constant because the secondary body’s orbital parameters change on a timescale much longer than its orbital period. Accordingly, on the timescale of a few orbital cycles, the trajectory admits a geodesic approximation, and we characterize the gravitational waveforms generated by such periodic orbits over a full cycle. It is worth noting that the effect of gravitational radiation on the small object's trajectory is negligible over such a short duration. To examine the gravitational waveforms, we begin by solving the equations of motion numerically, after which the waveforms are generated via the symmetric and trace-free (STF) mass quadrupole formula for gravitational radiation~\cite{Yang:2024lmj}. One can write it as~\cite{RevMod}
\begin{equation}
 I^{ij}=\left[\int d^3xx^ix^{j}T^{tt}(t,x^i)\right]^{\text{(STF)}}\,,
\end{equation}
with $T^{tt}$ serves as $tt$-component of the energy-momentum tensor for the secondary body on the trajectory $Z^i(t)$. It can be read as
\begin{equation}
    T^{tt}(t,x^i)=m\delta^3(x^i-Z^i(t))\,.
\end{equation}
It is essential to express the trajectory of the small celestial object in the Cartesian coordinate system by treating the Boyer-Lindquist coordinates as a fictitious spherical polar coordinate system
\begin{align}
x=r\sin{\theta}\cos{\phi}, \quad y=r\sin{\theta}\sin{\phi}, \quad z=r\cos{\theta}\, .
\end{align}
The metric perturbations can be defined as follows by taking into account the above equations
\begin{equation}
h_{ij}=\frac{2}{D_L}\frac{d^2I_{ij}}{dt^2}=\frac{2m}{D_L}(a_ix_j+a_jx_i+2v_iv_j)\,.
\end{equation}
The quantities $v_i$ and $a_i$ correspond, respectively, to the velocity and acceleration of the secondary celestial body. This waveform presented here reflects the fundamental manner in which GWs propagate through space. To ensure consistency with actual detector observations, the modeled GW signal must be transformed into the detector's reference frame. Accordingly, the GW must be projected onto a coordinate system aligned with the detector's orientation~\cite{Poisson_Will_2014, 2024arXiv241101858M}. One can therefore write the coordinate basis as
\begin{equation}\label{coordsystem}
\begin{array}{l}
e_{X}=[\cos\zeta,-\sin\zeta,0]\,,\\
e_{Y}=[\cos\iota\sin\zeta,\cos\iota\cos\zeta,-\sin\iota]\,,\\
e_{Z}=[\sin\iota\sin\zeta,\sin\iota\cos\zeta,\cos\iota]\,.
\end{array}
\end{equation}
Here, $\zeta$ designates the periastron longitude and $\iota$ represents the inclination angle of the test particle's orbit with respect to the line of sight. The orthogonal coordinate basis of the detector's reference frame is given by the vectors $e_X$, $e_Y$ and $e_Z$, which allow the GW signal to be decomposed into its separate polarization components. Hence, the associated GW polarizations are expressed as~\cite{Poisson_Will_2014}
\begin{eqnarray}
 h_{+}&=&\frac{1}{2}(e^i_X e^j_X-e^i_Ye^j_Y)h_{ij}, \\
 h_{\times}&=&\frac{1}{2}(e^i_Xe^j_Y+e^i_Ye^j_X)h_{ij}.
\end{eqnarray}
The above equation takes the following form when we consider Eq.~(\ref{coordsystem})
\begin{eqnarray}
h_{+}&=&\frac{1}{2}(h_{\zeta\zeta}-h_{\iota\iota}), \\
h_{\times}&=& h_{\iota\zeta}.
\end{eqnarray}
with
\begin{eqnarray}
h_{\zeta\zeta}&=&h_{xx}\cos^2{\zeta}-h_{xy}\sin{2\zeta}+h_{yy}\sin^2{\zeta}, \\
h_{\iota\iota}&=&\cos{\iota}[h_{xx}\sin^2{\zeta}+h_{xy}\sin{2\zeta}+h_{yy}\cos^2{\zeta}]\nonumber\\&+&h_{zz}\sin^2{\iota}-\sin{2\iota}[h_{xz}\sin{\zeta}+h_{yz}\cos{\zeta}], \\
h_{\iota\zeta}&=&\cos{\iota}\Big[\frac{1}{2}h_{xx}\sin{2\zeta}+h_{xy}\cos{2\zeta}-\frac{1}{2}h_{yy}\sin{2\zeta}\Big]\nonumber\\&+&\sin{\iota}[h_{yz}\sin{\zeta}-h_{xz}\cos{\zeta}].
\end{eqnarray}
To visualize the GWs, we consider the masses of the supermassive BH and stellar-mass object to be $M\sim 10^7 M_{\odot}$ and $m\sim 10M_{\odot}$, respectively. In addition, we set the inclination angle and longitude as $\iota=\zeta=\pi/4$, and the luminosity distance as $D_L=200 \text{Mpc}$. In Fig.~\ref{fig:comb}, we plot the $(3,1,2)$ periodic orbits for $\xi=0$ (red line) and $\xi=0.5$ (blue line) together with associated gravitational waveforms. This figure shows that gravitational waveforms include both calm and rapidly oscillating components, which, respectively, refer to the zoom and whirl stages. In addition, we can see that there is a slight shrink in the periodic orbits with the increase of the $\xi$ parameter. As an outcome, this causes the GW signal to arrive earlier. Moreover, we examine the individual parts of the periodic orbit $(1,2,0)$ displayed in Fig.~\ref{fig:120}. This figure thus clarifies which part of the periodic orbit maps onto which part of the gravitational waveforms. To provide further insight, we present the comparison between various periodic orbits in Fig.~\ref{fig:comp}. It should be emphasized that the carrying analysis is restricted to a single orbital period so that each orbit yields distinct gravitational waveforms.

\section{Conclusions}\label{sum}

In the present work, we have investigated the dynamics of massive particles and GW signatures of periodic orbits around a new improved Schwarzschild BH. We first examined the BH metric and explored how the cutoff parameter $\xi$ and interpolation parameter $\gamma$ influence the spacetime geometry together with horizon structure. We then derived the effective potential governing massive particle motion and investigated its dependence on the orbital angular momentum and spacetime parameters. Our numerical results indicate that the effective potential increases under the influence of the $\xi$ parameter and orbital angular momentum, while the interpolation parameter produces a comparatively modest rise. We subsequently determined the MBOs and ISCOs using the corresponding conditions. The results demonstrate that characteristic radii and angular momenta of these orbits decrease as the spacetime parameters increase, distinguishing the orbital properties of the new improved Schwarzschild BH from those of the standard Schwarzschild solution. 

We next analyzed periodic orbits characterized by three integers, i.e., ($z,w,v$). By examining the rational ratio of the azimuthal and radial frequencies, we established the conditions for identifying different periodic orbit configurations. The associated particle energies were obtained numerically using the bisection method for a representative choice of orbital angular momentum. The results indicate that increasing the zoom number $z$ produces more intricate orbital trajectories, whereas increasing the whirl number leads to additional loops near the BH before the particle reaches its subsequent apoapsis. These distinct orbital patterns provide a useful framework for investigating how the underlying spacetime geometry influences the motion of massive particles. 

Finally, we calculated the GW polarizations $h_+$ and $h_{\times}$ generated by an EMRI system composed of a stellar-mass object orbiting a supermassive new improved Schwarzschild BH. Using the numerical trajectories and symmetric trace-free mass quadrupole formalism, we constructed the gravitational waveforms associated with several representative periodic orbits. The waveforms exhibit both slowly varying and rapidly oscillating features, corresponding to the zoom and whirl portions of the orbital motion, respectively. A comparison between the standard Schwarzschild case and the improved spacetime with a non-zero cutoff parameter reveals that the periodic trajectories become slightly more compact as $\xi$ increases, leading to an earlier arrival of the corresponding GW signal. We also compared the waveforms generated by different periodic orbit configurations and demonstrated how variations in orbital structure are reflected in the temporal behavior of the two GW polarizations. 

Overall, our findings establish a connection between modifications of the Schwarzschild spacetime, the dynamics of periodic orbits, and their associated GW signatures. In particular, the dependence of orbital radii, trajectory morphology, and waveform timing on the parameters $\xi$ and $\gamma$ suggests that GW observations of EMRIs may provide a means of testing deviations from the standard Schwarzschild geometry. Further investigations incorporating radiation reaction effects and detector-specific sensitivity analysis will be important for assessing the observational distinguishability of these signatures and the prospects for constraining the spacetime parameters with future space-borne GW detectors.

\section*{ACKNOWLEDGEMENT}

This research was funded by the National Natural Science Foundation of China (NSFC) under Grant No. U2541210.

\bibliographystyle{apsrev4-1}  

\bibliography{ref}

\end{document}